\documentclass{jfm}
\usepackage{graphicx}
\usepackage{xcolor}
\usepackage{newtxtext}
\usepackage{newtxmath}
\usepackage{natbib}
\usepackage[colorlinks=true,citecolor=blue,linkcolor=blue,urlcolor=blue]{hyperref}

\usepackage{nicematrix}
\usepackage[most]{tcolorbox}
\usepackage{booktabs}
\usepackage{float}

\newtcolorbox[blend into=figures]{myfigure}[2][]{float=htb,
title={#2},#1}

\definecolor{mygreen}{RGB}{65,171,93}

\newcommand{\RomanNumeralCaps}[1]
\usepackage[french]{babel} %Man­ages cul­tur­ally-de­ter­mined ty­po­graph­i­cal (and other) rules, and hy­phen­ation pat­terns for a wide range of lan­guages. A doc­u­ment may se­lect a sin­gle lan­guage to be sup­ported, or it may se­lect sev­eral, in which case the doc­u­ment may switch from one lan­guage to an­other in a va­ri­ety of ways.
\usepackage[utf8]{inputenx} %Demande que les caractères accentués soient reconnus comme de l'utf8.
\usepackage[T1]{fontenc} %Demande que soient utilisées les fontes de LaTeX incluant les accents. (je ne sais pas si c'est compatible Mac et PC).
\usepackage{nicematrix}

\title{Assessment of coupled bilayer-cytoskeleton modelling strategy for red blood cell dynamics in flow}

\author{V. Puthumana\aff{1},
P. G. Chen\aff{1},
M. Leonetti\aff{2}\corresp{\email{marc.leonetti@univ-amu.fr}}, R. Lasserre\aff{3}\corresp{\email{remi.lasserre@inserm.fr}}, 
and M. Jaeger\aff{1}\corresp{\email{marc.jaeger@centrale-marseille.fr}}}

\affiliation{\aff{1}Aix Marseille Univ, CNRS, Centrale Marseille, M2P2, Turing Centre for Living systems, France
\aff{2}Aix Marseille Univ, CNRS, CINaM, Turing Centre for Living systems, Marseille, France
\aff{3}Aix Marseille Univ, CNRS, INSERM, CIML, Turing Centre for Living systems, Marseille, France}

\begin{document}

%%%%%%%%
\maketitle

%%%%%%%%
\begin{abstract}
The red blood cell (RBC) membrane is composed of a lipid bilayer and a cytoskeleton interconnected by protein junction complexes, allowing for potential sliding between the lipid bilayer and the cytoskeleton. Despite this biological reality, it is most often modelled as a single-layer model, a hyperelastic capsule or a fluid vesicle. Another approach involves incorporating the membrane's composite structure using double layers, where one layer represents the lipid bilayer and the other represents the cytoskeleton. In this paper, we computationally assess the various modelling strategies by analysing RBC behaviour in extensional flow and four distinct regimes that simulate RBC dynamics in shear flow. The proposed double-layer strategies, such as the vesicle-capsule and capsule-capsule models, account for the fluidity and surface incompressibility of the lipid bilayer in different ways. Our findings demonstrate that introducing sliding between the layers offers the cytoskeleton a considerable degree of freedom to alleviate its elastic stresses, resulting in a significant increase in RBC elongation. Surprisingly, our study reveals that the membrane modelling strategy for RBCs holds greater importance than the choice of the cytoskeleton's reference shape. These results highlight the inadequacy of considering mechanical properties alone and emphasise the need for careful integration of these properties. Furthermore, our findings fortuitously uncover a novel indicator for determining the appropriate stress-free shape of the cytoskeleton.

\end{abstract}

%%%%%%%%
%\begin{keywords}
%Authors should not enter keywords on the manuscript, as these must be chosen by the author during the online suboldsymbolission process and will then be added during the typesetting process (see \href{https://www.cambridge.org/core/journals/journal-of-fluid-mechanics/information/list-of-keywords}{Keyword PDF} for the full list).  Other classifications will be added at the same time.
%\end{keywords}

%%%%%%%%
\section{Introduction}
\label{Introduction}

The red blood cell (RBC) is a unique cell that, during maturation, frees all typical internal organisation \citep{mohandas_red_2008, moschandreou_measurement_2012}.
Its impressive mechanical performance is due to its membrane made up of a lipid bilayer supported by a cortical spectrin cytoskeleton \citep{evans_mechanics_1980, mohandas_mechanical_1994, mohandas_red_2008, moschandreou_measurement_2012}. This organisation gives the RBC a richness of dynamics in flow, critical to understanding blood rheology. Numerical modelling of the RBC presents a challenge in identifying the dominant features necessary to reproduce this richness. 

Some models retain the main membrane organisation into two structures of different nature, likely to slide in relation to each other \citep{krishnaswamy_cosserat-type_1996, noguchi_shape_2005, mcwhirter_deformation_2011, peng_multiscale_2010, peng_multiscale_2011, peng_deformation_2013, peng_lipid_2013, li_probing_2014, peng_erythrocyte_2014, peng_stability_2015, salehyar_deformation_2016, pivkin_biomechanics_2016, chang_mddpd_2016, chang_modelling_2017, zhu_prospects_2017, lu_boundary_2019}. Each structure then has its own mesh, allowing for different kinematics to be considered. However, although these models have provided compelling arguments in their favour \citep{zhang_multiple_2015, peng_mesoscale_2015, andreoni_continuum-_2018}, it is not the most common approach.

In the quest for simplicity, a single-layer model is often preferred, despite the inevitable compromise of sacrificing some mechanical properties. When modelling the lipid bilayer as a vesicle, it becomes challenging to consider the shear elasticity provided by the cytoskeleton, as the focus shifts to favouring the fluidity and incompressibility of the bilayer. Conversely, when modelling the cytoskeleton as a capsule, the fluid nature of the lipid bilayer is compromised in favour of the elasticity of the cytoskeleton. In any case, a single-layer model imposes the same kinematics on the cytoskeleton and the lipid bilayer. For the out-of-plane kinematics, this assumption is limited only to the rare cases of cytoskeleton detachment \citep{peng_lipid_2013, zhu_prospects_2017}. For the in-plane kinematics, the limitation of prohibiting the sliding of the cytoskeleton is less clear. The lipid bilayer, being in direct contact with the external environment, is susceptible to hydrodynamic stress and can, in turn, only drive the cytoskeleton via frictional forces that the lipid medium exerts on the junction proteins. Although the theoretical analysis of \cite{fischer_is_1992} suggests that the kinematics of the cytoskeleton and the lipid bilayer are identical in the tank-treading regime in shear flow, and the surface incompressibility constraint is transmitted from the lipid bilayer to the cytoskeleton, these findings cannot be generalised due to the lack of research in this area.

Vesicle modelling has proven useful in understanding the dynamics of flowing RBCs and is still used to study ensemble behaviour and its extrapolation to the rheology of blood \citep{brust_plasma_2014, kabacaoglu_sorting_2019, lu_scalable_2019}. However, the fundamental contribution of shear elasticity provided by the cytoskeleton cannot be overlooked \citep{mendez_-plane_2018, hoore_effect_2018}. To account for this, the  single-layer capsule representation has emerged as the more prevalent modelling strategy. 

While the various modelling strategies for RBCs contribute to our understanding of their dynamics, a comparative study to assess the impact of each simplification has not yet been undertaken, to our knowledge. Only the choice between a capsule represented by a network of springs or a continuum model has been considered \citep{omori_comparison_2011, tsubota_short_2014}. However, this may lead to bias in the identification of mechanical properties necessary to reproduce experimental observations. To be meaningful, a comparative study must minimise the specificities of numerical implementation for each choice. This entails treating the method of flow solution and the geometric representation of surfaces with the same precision. The numerical platform we have developed to investigate the fluid-structure interaction problem of surfactant-coated drops, vesicles and capsules, enables such study \citep{boedec_isogeometric_2017}. In the vesicle model, we employ distinct approaches to handle the movements of the bilayer in the normal and tangential directions. Specifically, we adopt a Lagrangian approach for the normal movement, while we use an Eulerian description for the tangential movement. Consequently, the tangential motion of mesh vertices (or nodes within a finite element framework), which doesn't alter the membrane's shape, is entirely decoupled from the tangential movement of the lipids. In fact, this decoupling allows us to conveniently prescribe the tangential velocities of mesh vertices to maintain mesh quality in the context of vesicle simulations. Additionally, in \cite{lyu_hybrid_2018}, we demonstrated that different in-plane kinematics of the cytoskeleton and lipid bilayer can be considered without duplicating meshes. This is achieved by using an Eulerian kinematic description for the fluid lipid bilayer and a Lagrangian one for the solid cytoskeleton. The motion of the mesh coincides with that of the cytoskeleton, while the Eulerian description allows for different tangential motions of the lipid bilayer. Using only one mesh prohibits consideration of different out-of-plane kinematics but allows different in-plane ones at a minimal cost.

We believe that a comparative study to characterise the impact of different RBC modelling strategies is justified for another important reason: the contribution of surface viscosity. The origin of this viscosity can be explained by several mechanisms, but their inclusion in simulations depends on the choice of membrane representation~\citep{fedosov_multiscale_2014, li_continuum-_2013, vlahovska_flow_2013, yazdani_influence_2013, freund_numerical_2014, gounley_computational_2015, peng_mesoscale_2015, prado_viscoelastic_2015, andreoni_continuum-_2018, guglietta_effects_2020, tsubota_elongation_2021}. For instance, the friction of the junction proteins in the lipid bilayer can lead to significant variation in the effective surface viscosity of the RBC. Therefore, we feel it is essential to first assess the purely elastic aspects of RBC mechanics before considering the contribution of surface viscosity.

The rest of this paper is organised as follows. In \S\ref{sec:modelling}, we outline the implementation of the different modelling strategies, which are then subjected to a comparative study of their elastic properties under axisymmetric extensional flow in \S\ref{ElongationFlow} and simple shear flow in \S\ref{ShearFlow}, respectively. The former, akin to laser tweezer stretching, is a widely-used technique for characterising the mechanical properties of cells, while the latter is the reference flow for characterising the dynamics of RBCs. In \S\ref{sec_Discussion}, we discuss the results and their implications and provide new insights into the dynamics of RBCs. Finally, in \S\ref{sec_Conclusion}, we draw broader conclusions that extend beyond the modelling framework.

\section{RBC modelling strategies} \label{sec:modelling}

We have implemented the vesicle and capsule models in our platform, which have been previously described in \citet{boedec_isogeometric_2017} and \citet{lyu_isogeometric_2021} for use in unbounded and confined spaces, respectively. The vesicle model accounts for the fluid nature of the membrane and enforces surface incompressibility by projecting the 3D velocity field onto a space with zero surface divergence \citep{boedec_linear_2011}. In addition, the isogeoparametric representation of the geometry \citep{cottrell_isogeometric_2009}, which has at least $C^1$ regularity, allows for the incorporation of the Helfrich bending energy using a rigorous weak mathematical formulation. Meanwhile, the capsule model can utilise several models of polymerized membrane behaviour and incorporate bending elasticity through the Helfrich formulation developed for the vesicle model or other common forms of thin-shell modelling. Although the vesicle and capsule models can incorporate shear and dilatation surface viscosity developed for a surfactant drop \citep{gounley_influence_2016}, we will not be utilising this option for our objective of this paper, which is to assess the purely elastic aspects of RBC mechanics. As the details of these models are available elsewhere, we will focus solely on their key elements in this section.

Our isogeometric representation is obtained using the Loop subdivision method \citep{loop_smooth_1987, cirak_subdivision_2000, cirak_fully_2001}. To generate the control mesh for a vesicle or capsule, we start with an icosahedron and perform $N$ Loop subdivisions. The value $N=0$ corresponds to the initial icosahedron, which has $N_{elem}=20$ equilateral triangular elements that correspond to 20 Loop elements. With each refinement level, each element is subdivided into four equivalent equilateral triangles, and a node is inserted at the center of each edge. The spatial position of all nodes in the refined mesh is determined using Loop's rules, which ensure zero geometric representation error. The level $N$ generates a mesh with $N_{elem}=20 \times 4^N$. We typically achieve satisfactory accuracy using $N=3$, resulting in $N_{elem}=1280$. Note that the twelve nodes of the $N=0$ level have fifth-order connectivity, while all other nodes introduced afterwards have sixth-order connectivity. These twelve nodes are called singular, and they are an inevitable trace of the icosahedron. The regularity of the Loop approximation is only $C^1$ at these nodes, while it is $C^2$ everywhere else. It's worth highlighting that our isogeometric finite-element method, which relies on the Loop subdivision, diverges from traditional shell modelling techniques commonly used in structural mechanics. The advantage of our approach is its ability to circumvent the shear-locking phenomenon typically observed in conventional methods~\citep{cirak_subdivision_2000, cirak_fully_2001}.

\begin{figure}
\centerline{\includegraphics[width=5in]{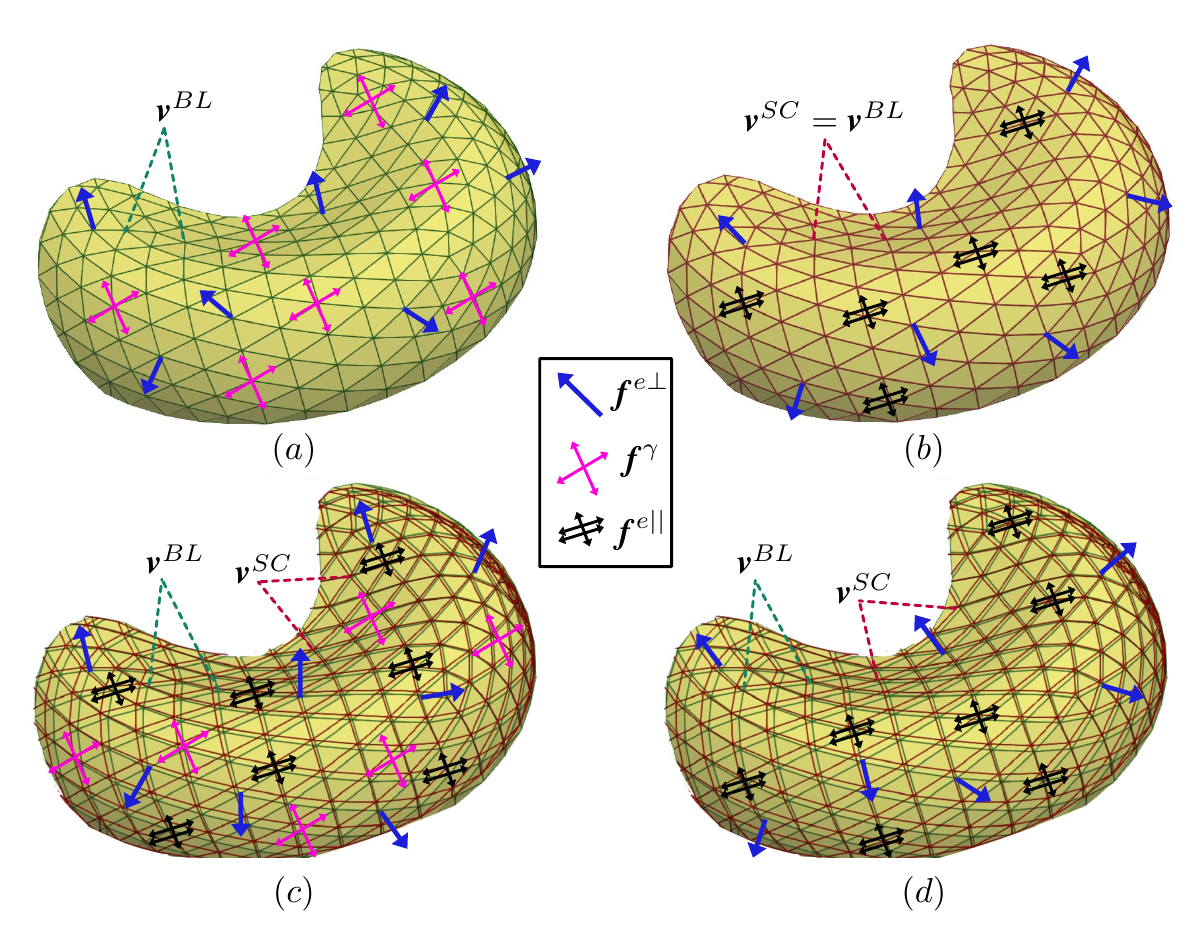}}
\caption{Schematic representation of the RBC models illustrating the relevant surface force densities on the respective layers.
(\textit{a}) Single-layer vesicle model with out-of-plane elasticity ($\boldsymbol{f}^{e\perp}$) and surface incompressibility ($\boldsymbol{f}^{\gamma}$).
(\textit{b}) Single-layer capsule model with in-plane ($\boldsymbol{f}^{e\parallel}$) and out-of-plane elasticities (surface incompressibility is approximatively considered).
(\textit{c}) Double-layer vesicle-capsule model with in-plane and out-of-plane elasticities, along with surface incompressibility.
(\textit{d}) Double-layer capsule-capsule model with in-plane and out-of-plane elasticities (surface incompressibility is approximatively considered).
All models employ a single mesh, with the distinction that in double-layer models, $\boldsymbol{v}^{BL} \neq \boldsymbol{v}^{SC}$ due to the differing assumed tangential kinematics of the two layers.}
\label{RBC_models}
\end{figure}

The boundary element method \citep{pozrikidis_boundary_1992, pozrikidis_practical_2002} is used with the surface mesh described earlier to obtain the velocity of the lipid medium at each node. We can then determine the velocity at any point of an element using Loop's interpolation functions. The normal component of this velocity field determines the evolution of the surface position between two time steps in the simulation. Meanwhile, the movement of the nodes tangentially to the membrane is identified with the kinematics of a cytoskeleton. This tangential movement is determined by equilibrating the elastic forces exerted by the cytoskeleton with the viscous friction forces of the points of attachment of the cytoskeleton in the lipid bilayer. The viscous friction force is proportional to the velocity difference between the bilayer and the cytoskeleton. For the unconfined case and without considering a viscosity contrast to simplify the presentation, the system can be formulated as follows \citep{lyu_hybrid_2018}:

\begin{equation}
\label{BaseEquationForV}
\boldsymbol{v}^{BL} = \boldsymbol{v}^{3D} = \mathsfbi{P} \left( \boldsymbol{v}^{\infty} + \mathsfbi{G} \boldsymbol{f}^{RBC} \right) 
\quad \mbox{with} \quad
\boldsymbol{f}^{RBC} = \boldsymbol{f}^{BL}  + \boldsymbol{f}^{SC},
\end{equation}
\begin{equation}
\label{BaseEquationForX}
\boldsymbol{x}(t+\mathrm{d}t) = \boldsymbol{x}(t) + \boldsymbol{v}^{SC}\mathrm{d}t
\quad \mbox{with} \quad
\boldsymbol{v}^{SC} = \boldsymbol{v}^{BL} + \boldsymbol{v}^{SC/BL}.
\end{equation}

Position and velocity fields of the RBC membrane are denoted by $\boldsymbol{x}$ and $\boldsymbol{v}$, respectively. Superscripts BL and SC are used to distinguish between the lipid bilayer and the spectrin cytoskeleton, respectively, while $3D$ is for the bulk flow and $\infty$ for the imposed background one. 
The velocity difference between the cytoskeleton and the lipid bilayer $ \boldsymbol{v}^{SC/BL} = \boldsymbol{v}^{SC} - \boldsymbol{v}^{BL} $ is a tangential vector.
The Green's operator associated with the Stokeslet is represented by $\mathsfbi{G}$, and the projection operator on a subspace with zero surface divergence is denoted by $\mathsfbi{P}$. This operator ensures that the surface tension $\gamma$, which is the Lagrange multiplier of the surface incompressibility constraint, satisfies the condition that the surface divergence of the velocity field $\boldsymbol{v}^{BL}$ is zero. The surface densities of force induced by the lipid bilayer and the cytoskeleton correspond to $\boldsymbol{f}^{BL}$ and $\boldsymbol{f}^{SC}$, respectively, and their sum $\boldsymbol{f}^{RBC}$ corresponds to the surface force density exerted by the entire RBC membrane on the ambient fluids. The matrix expression of the system (\ref{BaseEquationForV})--(\ref{BaseEquationForX}) and the principle of the solution algorithms are presented in detail in \cite{boedec_isogeometric_2017} and \cite{lyu_hybrid_2018, lyu_isogeometric_2021}. While we encourage interested readers to refer to these sources, we note that the rest of the paper is accessible without prior knowledge of them. 

Using this validated approach, which has been successfully applied in our previous studies of vesicles and capsules, we have developed several strategies for modelling RBCs. These strategies encompass single-layer vesicle and capsule models, as well as double-layer vesicle-capsule and capsule-capsule models. To aid in visual comprehension, figure~\ref{RBC_models} presents a schematic depiction of the surface densities relevant to the RBC models. In the subsequent sections, we provide a detailed outline of the implementation for each of these models.
To aid in the comparison of the results across all the studies, the same colour coding is utilised in all figures, where the vesicle is represented in black, the capsule in blue, the vesicle-capsule in red, and the capsule-capsule in green.

Throughout  this paper, we use dimensionless variables, denoted by a star symbol. Lengths are expressed in units of a volume-based radius $R\equiv [3V/(4\upi)]^{1/3}$, where $V\simeq94$~$\upmu$m$^3$ represents the enclosed volume of RBCs.  The surface area $A$ is about $135~\upmu$m$^2$, giving a reduced volume $v\equiv3\sqrt{4\pi}VA^{-3/2} = 0.64$~\citep{evans_mechanics_1980}. Time is scaled by $\eta_{ext}R/\mu_s$, where $\eta_{ext}$ is the viscosity of the suspending fluid and $\mu_s$ is the surface shear modulus. A summary of the mechanical properties utilised in the simulations is provided in table~\ref{RBC_Properties_Elasticity}. These data are based on averages of recognised values for a healthy RBC [see, e.g. table 1 in \cite{levant_intermediate_2016}]. The table also displays the distribution of properties between the cytoskeleton and the lipid bilayer in the modelling strategies that distinguish them.

The reference shape, whether quasi-spherical or biconcave, represents the shape relative to which in-plane and out-of-plane deformations are defined. In simpler terms, when the cytoskeleton (Skalak capsule) adopts the reference shape, the in-plane deformations are zero.  It's important to clarify that all our simulations begin with an RBC already in the biconcave shape (i.e. $v=0.64$), which we obtain directly from an analytical expression~\citep{evans_mechanics_1980} and consistently used in our prior work, as outlined in~\citet{lyu_hybrid_2018}. The reference shapes we consider include the biconcave shape itself, a quasi-spherical shape ($v=0.96$) with zero spontaneous curvature, and a quasi-spherical shape ($v=0.96$) with positive curvature ($C_0^* = C_0/R = 4$)~\citep{peng_erythrocyte_2014}. If not specified, we have used quasi-spherical with zero spontaneous curvature. The quasi-spherical reference shape, achieved through a relaxation process, corresponds to the deflated shape of a vesicle with a reduced volume of $v=0.96$.

%In our study, we primarily adopted the quasi-spherical shape with a sphericity of 0.96, which is a widely used choice in the numerical modelling of RBCs. We intentionally avoided incorporating any spontaneous curvature into our models. However, to examine the sensitivity of our results to this selection, we occasionally employed the discoidal shape of the resting RBC and introduced a spontaneous curvature ($C_0^* = C_0/R = 4$) for the quasi-spherical shape \citep{peng_erythrocyte_2014}. This analysis is discussed in relation to each of the considered flow configurations.

\begin{table}
\begin{center}
\def~{\hphantom{0}}
  \begin{tabular}{ccccccc}
    Property (units) & Vesicle & \textcolor{blue}{Capsule} & \textcolor{red}{Vesicle-} & \textcolor{red}{capsule} & \textcolor{mygreen}{Capsule-} & \textcolor{mygreen}{capsule} \\[3pt]
    In-plane elasticity (\ref{W_Capsule}) & & & & & & \\
    $ G_s$ ($\upmu$N m$^{-1}$)  & --- & 6.0 & --- & 6.0 & $10^{-3}$ & 6.0 \\
    $ C^{SK}$ | $k_s$ ($\upmu$N m$^{-1}$)  & --- & $ 80.0 \, | \, 10^{3} $ & --- & $ 2.0 \, | \, - $ & $ 80.0 \, | \, 10^{3} $ & $ 2.0 \, | \, - $ \\
    \\
    Out-of-plane elasticity (\ref{W_Vesicle}) & & & & & & \\
    $k_b$ ($\times 10^{-19}$ J)  & 2.4 & 2.4 & 2.4 & --- & 2.4 & --- \\
    \\
    Cytoskeleton/bilayer friction (\ref{Cf}) & & & & & & \\
    $ C_f$ (pN s $\upmu$m$^{-3}$)  & --- & --- & --- & 144.0 & ---  &144.0 \\
    \\
     Inner viscosity & & & & & & \\
    $ \eta_{int}$ (mPa s) & 10.0 & 10.0 & --- & 10.0 & --- & 10.0 \\
  \end{tabular}
  \caption{Mechanical properties.}
  \label{RBC_Properties_Elasticity}
   \end{center}
\end{table}

\subsection{Single-layer vesicle strategy (colour code = black)}
\label{sec:vesicle}

The vesicle model doesn't consider a separate contribution of the cytoskeleton (i.e. $ \boldsymbol{f}^{SC} = \boldsymbol{0} $), while

\begin{equation}
\label{fBL_Vesicle}
\boldsymbol{f}^{BL}(\boldsymbol{x}) = \boldsymbol{f}^{\gamma}(\boldsymbol{x}) + \boldsymbol{f}^{e\perp}(\boldsymbol{x}) 
= -\frac{\delta \mathcal{W}^{BL}}{\delta \boldsymbol{x}} 
= -\frac{\delta ( \mathcal{W}^{\gamma} + \mathcal{W}^{H} )}{\delta \boldsymbol{x}} ,
\end{equation}
\begin{equation}
\label{W_Vesicle}
\mathcal{W}^{\gamma} = \int_S \gamma \:\mathrm{d} S,  \quad  \mathcal{W}^{H} =  \int_S w^H \:\mathrm{d}S, \quad 
w^H = \frac{k_b}{2} (2 H + C_0 )^2.
\end{equation}
The $\boldsymbol{f}^{e\perp}$ contribution represents the out-of-plane elastic forces, namely the bending forces induced by the Helfrich surface energy density $w^H$, which depend on the mean curvature $H$ and the spontaneous curvature $C_0$. The spontaneous curvature $C_0$ is defined as $ C_0 = - 2H_0 $, where $H_0$ is the mean curvature of the spontaneous or reference shape of the lipid bilayer. For zero spontaneous curvature, the Helfrich energy reduces to the simple expression $ w^H = 2 k_b H^2 $, where $k_b=2.4\times10^{-19}$ J is the bending modulus. We omit the contribution of the Gaussian curvature, which can occur in $w^H$ but is not useful due to the Gauss-Bonnet theorem, as we are not considering a change in surface topology. The contribution of $\boldsymbol{f}^{\gamma}$ corresponds to the forces induced by the surface tension $\gamma$. For a constant tension, it results in an out-of-plane force contribution, but it can also produce in-plane forces if the tension varies, as $ \boldsymbol{f}^{\gamma} = 2 \gamma H \boldsymbol{n} + \bnabla_s \gamma $, where $\boldsymbol{n}$ is the normal vector to the surface and $\bnabla_s= (\mathsfbi{I} - \boldsymbol{n} \boldsymbol{n}) \bcdot \bnabla$ is the surface gradient operator, with $\mathsfbi{I}$ being the identity in $\mathbb{R}^3$.

We want to emphasise that in the vesicle model, surface incompressibility, expressed as $\bnabla_s \bcdot \boldsymbol{v}^{BL} = 0$, is enforced through a projection algorithm, specifically, the projection operator $\mathsfbi{P}$ as defined in eq.~(\ref{BaseEquationForV}). This projection introduces the surface tension parameter $\gamma$, acting as a Lagrange multiplier for the surface incompressibility constraint. While this method effectively ensures both local and global surface area conservation, it does come at a significant computational cost, as previously discussed~\citep{boedec_isogeometric_2017}. 

\subsection{Single-layer capsule strategy (colour code = \textcolor{blue}{blue})}
\label{sec:capsule}

The capsule model doesn't consider a separate contribution of the cytoskeleton (i.e. $ \boldsymbol{f}^{SC} =  \boldsymbol{0} $). While bending is still modelled using the Helfrich energy, the in-plane elasticity of the cytoskeleton is nicely represented by a capsule. On the other hand, fluidity is lost and surface incompressibility can only be approximated since the projector $\mathsfbi{P}$ is replaced by the identity. Thus, we obtain the capsule model with

\begin{equation}
\label{fBL_Capsule}
\boldsymbol{f}^{BL}(\boldsymbol{x}) = \boldsymbol{f}^{e \parallel}(\boldsymbol{x}) + \boldsymbol{f}^{e\perp}(\boldsymbol{x})
= -\frac{\delta \mathcal{W}^{BL}}{\delta \boldsymbol{x}}
= -\frac{\delta ( \mathcal{W}^{SK} + \mathcal{W}^{H} )}{\delta \boldsymbol{x}} ,
\end{equation}
where $\boldsymbol{f}^{e\parallel}$ represents the in-plane elastic forces deduced from a polymerized membrane strain energy, which is usually defined on the reference configuration ($S^0$). Surface deformations are computed relative to this reference configuration \citep[eq. 24]{boedec_isogeometric_2017}. To model the strain-hardening behaviour of the RBC membrane, we typically use the expression proposed by Skalak~\citep{skalak_strain_1973}, which is given by

\begin{equation}
\label{W_Capsule}
\mathcal{W}^{SK} = \int_{S^0} w^{SK} \:\mathrm{d} S^0, \quad
w^{SK} = \frac{G_s}{4} ( I_1^2 + 2I_1 - 2I_2 + C^{SK}I_2^2 ),
\end{equation}
where the two invariants $I_1$ and $I_2$ are expressed as functions of the two main strains $\lambda_1$ and $\lambda_2$. The resistance to shear is controlled by the coefficient $G_s$, which is the surface shear modulus $ \mu_s = 6$ $\upmu$N m$^{-1}$, and the resistance to local area variation is controlled by the coefficient $C^{SK}$, which is the ratio of the area dilatation modulus to the shear modulus. The limiting case of surface incompressibility is obtained by making $C^{SK}$ tend towards infinity, where $I_2$ cancels. In practice, however, the numerical problem becomes too steep when $C^{SK}$ exceeds a hundred~\citep{dodson_tank-treading_2010}. For long-term simulations, we add a global area conservation constraint to prevent a strong drift of the RBC surface area. This constraint results in a contribution of the same type as $\mathcal{W}^{\gamma}$, with the surface tension $\gamma$ being replaced by a constant tension $k_s (S - S^0)/S^0$. In all our simulations, we used $C^{SK} = 80$ in combination with $k_s = 10^{3}$ $\upmu$N\,m$^{-1}$\citep{siguenza_how_2017} when surface incompressibility has to be considered. Our extensive investigation has revealed that the contribution of $k_s$ remains negligible, as local area preservation is consistently maintained throughout our study.

\subsection{Double-layer vesicle-capsule strategy (colour code = \textcolor{red}{red})}
\label{sec:vesicle-capsule}

The vesicle-capsule model is obtained by combining the expression (\ref{fBL_Vesicle}) of $\boldsymbol{f}^{BL}$ and 

\begin{equation}
\label{fSC_Capsule}
\boldsymbol{f}^{SC}(\boldsymbol{x}) = \boldsymbol{f}^{e \parallel}(\boldsymbol{x}) 
= -\frac{\delta \mathcal{W}^{SC}}{\delta \boldsymbol{x}} 
= - \frac{\delta \mathcal{W}^{SK}}{\delta \boldsymbol{x}}.
\end{equation}
In contrast to the vesicle model, the contribution from the cytoskeleton $\boldsymbol{f}^{SC}$ is not set to zero. The incompressibility constraint is still rigorously ensured by the $\mathsfbi{P}$ projector. The coefficient $C^{SK}$ in Skalak's law is then fixed to represent the contribution of the cytoskeleton only, which is negligible compared to the resistance to expansion provided by the lipid bilayer. The dilatation modulus of the cytoskeleton was determined experimentally to be within the range of 1--10 $\upmu$N m$^{-1}$~\citep{lenormand_direct_2001, lenormand_elasticity_2003, moschandreou_measurement_2012}, with a theoretical value for a regular hexagonal lattice being twice the shear modulus~\citep{discher_molecular_1994, hansen_elastic_1996}. Thus, we set the expansion rigidity of the capsule representing the cytoskeleton to $C^{SK} = 2$ in our double-layer strategy simulations. In other words, the cytoskeleton only provides resistance to shear deformation, with $ G_s = 6$ $\upmu$N m$^{-1}$ and $C^{SK}=2$.

\subsection{Double-layer capsule-capsule strategy (colour code = \textcolor{mygreen}{green})}
\label{sec:capsule-capsule}

The distinction between the vesicle-capsule model and the capsule-capsule model lies in how the lipid bilayer is represented, with the former using a vesicle  and the latter using a capsule to mimic an incompressible fluid membrane. This disparity is particularly noticeable in eq.~(\ref{fBL_Capsule}), which incorporates the necessary bending energy to represent the bilayer. In contrast, eq.~(\ref{fSC_Capsule}) lacks a bending energy term, which is essential for modelling the cytoskeleton. These variations in mechanical properties are outlined in detail in table~\ref{RBC_Properties_Elasticity}.

To derive the capsule-capsule model, we combine eq.~(\ref{fBL_Capsule}) for $\boldsymbol{f}^{BL}$ and eq.~(\ref{fSC_Capsule}) for $\boldsymbol{f}^{SC}$. Similar to the vesicle-capsule model, Skalak's law is used to represent the cytoskeleton, employing the same parameter values ($G_s = 6$ $\upmu$N m$^{-1}$ and $C^{SK} = 2$). However, achieving the behaviour of the lipid bilayer in $\boldsymbol{f}^{e \parallel}$ within $\boldsymbol{f}^{BL}$ necessitates another capsule model and a second application of Skalak's law. To accomplish this, the elastic shear modulus $G_s$ is chosen to be sufficiently small to ensure that the shear strength contribution remains negligible in comparison to the cytoskeleton. For this purpose, we have selected $G_s = 10^{-3}$ $\upmu$N m$^{-1}$ \citep{zhu_prospects_2017}, alongside $C^{SK} = 80$ and $k_s = 10^{3}$ $\upmu$N m$^{-1}$. While this approach does not perfectly emulate the fluid nature of the lipid bilayer due to the non-zero shear modulus, it enhances computational efficiency.

\subsection{Membrane elastic forces determination}
\label{sec:elastic-forces}

The boundary conditions between the two-layer structure are implicit. Because of the single mesh representation, the two structures are constraint to have the same motion in the normal direction. In the tangent plane, sliding is permitted and limited exclusively by tangential friction forces. Therefore, the double-layer models depend on the frictional coupling between the lipid bilayer and cytoskeleton through junction protein complexes. Let $\boldsymbol{f}^{BL/SC}$ be the surface density of frictional force exerted by the lipid bilayer on the cytoskeleton. By the action-reaction principle, $\boldsymbol{f}^{SC/BL} = -\boldsymbol{f}^{BL/SC}$. The static equilibrium of the two structures can be expressed as:
\begin{equation}
\boldsymbol{f}^{SC} + \boldsymbol{f}^{BL/SC} = 0, \quad \boldsymbol{f}^{BL} + \boldsymbol{f}^{SC/BL} + \boldsymbol{f}^{ext} = 0,
\end{equation}
where $\boldsymbol{f}^{ext}$ represents the action of the ambient fluids on the RBC membrane, which is in equilibrium with the forces induced by the deformation of the RBC, namely

\begin{equation}
\boldsymbol{f}^{ext} = -\boldsymbol{f}^{RBC} = -(\boldsymbol{f}^{BL} + \boldsymbol{f}^{SC}), 
\end{equation}
consistent with eq.~(\ref{BaseEquationForV}).

The finite element method is used to determine the external forces through a weak formulation using the virtual work principle, for a given membrane configuration

\begin{equation}
\delta \mathcal{W}^{ext} + \delta \mathcal{W}^{int} = 0,
\end{equation}

\begin{align}
\delta \mathcal{W}^{ext} 
&= \int_S \boldsymbol{f}^{ext} \bcdot \delta\boldsymbol{x}\:\mathrm{d}S 
= - \int_S \boldsymbol{f}^{RBC} \bcdot \delta\boldsymbol{x}\:\mathrm{d}S ,
\\
\delta \mathcal{W}^{int} 
&= -( \delta \mathcal{W}^{BL} + \delta \mathcal{W}^{SC} ) 
= \int_S -\frac{\delta \mathcal{W}^{BL}}{\delta \boldsymbol{x}} \bcdot \delta\boldsymbol{x}\:\mathrm{d}S 
+ \int_S -\frac{\delta \mathcal{W}^{SC}}{\delta \boldsymbol{x}} \bcdot \delta\boldsymbol{x}\:\mathrm{d}S 
\notag \\
& = \int_S \boldsymbol{f}^{BL} \bcdot  \delta\boldsymbol{x}\:\mathrm{d}S 
+ \int_S \boldsymbol{f}^{SC} \bcdot  \delta\boldsymbol{x}\:\mathrm{d}S.
\end{align}

The internal virtual work $ \delta \mathcal{W}^{int} $ depends on the surface stress and strain tensors, which can be determined based on the membrane configuration, 
see \cite{boedec_isogeometric_2017} for details: 

 \begin{align}
\delta \mathcal{W}^{int} 
&= -\int_S \left[ \sigma^{\alpha\beta} \delta(E_{\alpha\beta}) + \mu^{\alpha\beta} \delta(B_{\alpha\beta})\right]\:\mathrm{d}S 
= \int_S \left[\frac{1}{2} \sigma^{\alpha\beta} \delta(a_{\alpha\beta})   + \mu^{\alpha\beta} \delta(b_{\alpha\beta})\right]\:\mathrm{d}S.
\end{align}
Here, $\sigma_{\alpha\beta}$ and $\mu_{\alpha\beta}$ correspond to the stresses induced by membrane and bending strains, respectively. The tensor of components $E_{\alpha\beta}$ with $ 2E_{\alpha\beta} = a_{\alpha\beta}-a^{0}_{\alpha\beta} $ is the surface Green-Lagrange strain tensor and the tensor of components
$B_{\alpha\beta} = b_{\alpha\beta}-b^{0}_{\alpha\beta} $  is the bending equivalent, where superscript 0 refers to the reference configuration. 
The $a_{\alpha\beta}$ are components of the metric tensor $\mathsfbi{a}$ that correspond to the identity operator in the tangent plane $\mathsfbi{I}^S = \mathsfbi{I} - \boldsymbol{n} \boldsymbol{n}$. 
Its determinant $a = \mathrm{det}(\mathsfbi{a})$ expresses the surface element as a function of the surface parameterization $(s^1, s^2)$ as $ \mathrm{d}S = \sqrt{a} \,\mathrm{d}s^1 \mathrm{d}s^2 $. 

\subsection{Cytoskeleton drag forces determination}
\label{sec:drag-forces}

We define $C_{f_{JC}}$ as the mean friction coefficient of a protein junction complex in the lipid bilayer. The frictional force generated by the movement of a single junction complex,  on average, can be expressed as 

\begin{align}
\boldsymbol{f}^{BL/SC}_{JC} 
&= C_{f_{JC}} \left( \boldsymbol{v}^{BL}(\boldsymbol{x}_{JC}) - \boldsymbol{v}^{SC}(\boldsymbol{x}_{JC})  \right) \notag \\
&= C_{f_{JC}} \boldsymbol{v}^{BL/SC}(\boldsymbol{x}_{JC}) 
= -C_{f_{JC}} \boldsymbol{v}^{SC/BL}(\boldsymbol{x}_{JC}).
\end{align}
The equivalent friction coefficient per unit area of the membrane, denoted as $C_{f}$, is given by 
\begin{equation}
\int_{\triangle S} C_{f} \boldsymbol{v}^{BL/SC}(\boldsymbol{x})\:\mathrm{d}S
= \sum_{\boldsymbol{x}_{JC} \in \triangle S} C_{f_{JC}} \boldsymbol{v}^{BL/SC}(\boldsymbol{x}_{JC}) .
\end{equation}
For a small patch of surface $\triangle S$, where the velocity variations can be ignored (i.e. $\forall \boldsymbol{x} \in \triangle S$, $\boldsymbol{v}^{BL/SC}(\boldsymbol{x}) = \boldsymbol{v}^{BL/SC}(\boldsymbol{x}_{JC}) \approx$ constant), we can simplify the expression to 

\begin{equation}
C_{f} \triangle S = C_{f_{JC}} N_{JC} ,
\end{equation}
 where $N_{JC}$ is the number of junction complexes within the patch. We can also introduce the areal density of the junction complex, denoted as $\rho_{JC}$, and obtain 
 
 \begin{equation}
\label{Cf}
C_{f} = \frac{N_{JC}}{\triangle S} C_{f_{JC}} = \rho_{JC} C_{f_{JC}}.
\end{equation}
For our study, we adopted $C_f = 144$ pN s $\upmu$m$^{-3}$~\citep{peng_multiscale_2011}. The sensitivity of the double-layer modelling strategies to this parameter is discussed in \S\ref{sec_Discussion}.

\subsection{Cytoskeleton kinematics}
\label{sec:kinematics}

The velocity differential between the cytoskeleton and the lipid bilayer is defined by the expression

\begin{equation}
\boldsymbol{v}^{SC/BL}(\boldsymbol{x})
= \boldsymbol{v}^{SC}(\boldsymbol{x}) - \boldsymbol{v}^{BL}(\boldsymbol{x})
= - \frac{1}{C_f} \boldsymbol{f}^{BL/SC}(\boldsymbol{x})
= \frac{1}{C_f} \mathsfbi{I}^S  \boldsymbol{f}^{SC}(\boldsymbol{x}).
\end{equation}
To obtain the weak formulation of the problem, we employ the weighted residual method with $\delta \boldsymbol{x}$ as the test function, leading to the expression

\begin{equation}
\int_S \boldsymbol{v}^{SC/BL}(\boldsymbol{x}) \bcdot \delta \boldsymbol{x}\:\mathrm{d}S
= \frac{1}{C_f} \int_S \mathsfbi{I}^S \boldsymbol{f}^{SC}(\boldsymbol{x}) \bcdot \delta \boldsymbol{x}\:\mathrm{d}S.
\end{equation}
The first member of this equation can be expressed as a function of the mass matrix in terms of the Loop interpolation functions, as given in \citet{boedec_isogeometric_2017}. Notably, the second term in the equation, which represents the internal force density of the cytoskeleton, is similar to the corresponding expression used to determine the elastic membrane forces of the RBC.

\section{Comparison of RBC modelling strategies on extensional flow}
\label{ElongationFlow}

Laser tweezer stretching is a widely used method to measure the mechanical properties of cells. However, numerical studies have two major shortcomings. Firstly, there is variability in how the opposing forces are applied in simulations~\citep{siguenza_how_2017}. Secondly, it is not representative of the stretch that a red blood cell may experience in a flow since it does not consider the interaction with the surrounding fluid. An alternative approach that avoids these limitations is extensional flow, which has been used for studying vesicles in previous studies \citep{kantsler_vesicle_2007, kantsler_critical_2008, zhao_shape_2013, narsimhan_mechanism_2014, dahl_experimental_2016}. In this method, the flow is defined by a single parameter, $\dot{\epsilon}$, which represents the stretching rate. In a Cartesian coordinate system, the velocity component in the direction of stretching ($z$ coordinate) is given by $ v_z = \dot{\epsilon} z $. In its planar version, the other two velocity components are $ v_x = - \dot{\epsilon} x $ and $ v_y = 0 $. In its axisymmetric version, they are written as $ v_x = - \dot{\epsilon} x /2 $ and $ v_y = - \dot{\epsilon} y /2 $. We considered both configurations, with the axis of symmetry of the RBC along the $y$ axis and a value of $\dot{\epsilon} = 55$ s$^{-1}$. The viscosity of the surrounding fluid was set at $ \eta_{ext} =25$ mPa s. As the behaviours were found to be similar, we only present the results for the axisymmetric configuration. 

We first checked that the double-layer models behave similarly to the single-layer capsule model when the layers are prevented from sliding. 
This similarity is expected because all models employ the same set of elastic properties, as outlined in table \ref{RBC_Properties_Elasticity}.
The difference lies in the distribution of these properties. In the capsule-capsule model, the outer capsule mimics the lipid bilayer and assumes responsibility for bending resistance and area conservation. The coefficient of shear elasticity for the outer capsule is considered negligible compared to that of the inner capsule, which solely represents the cytoskeleton properties. The vesicle-capsule model follows a similar approach. However, as the lipid bilayer is now represented by a vesicle, surface incompressibility is treated more rigorously. Consequently, a slightly reduced elongation compared to the other models is observed, reflecting the stiffening effect caused by the surface incompressibility constraint.

\begin{figure}
\begin{center}
\includegraphics[width=5.4in]{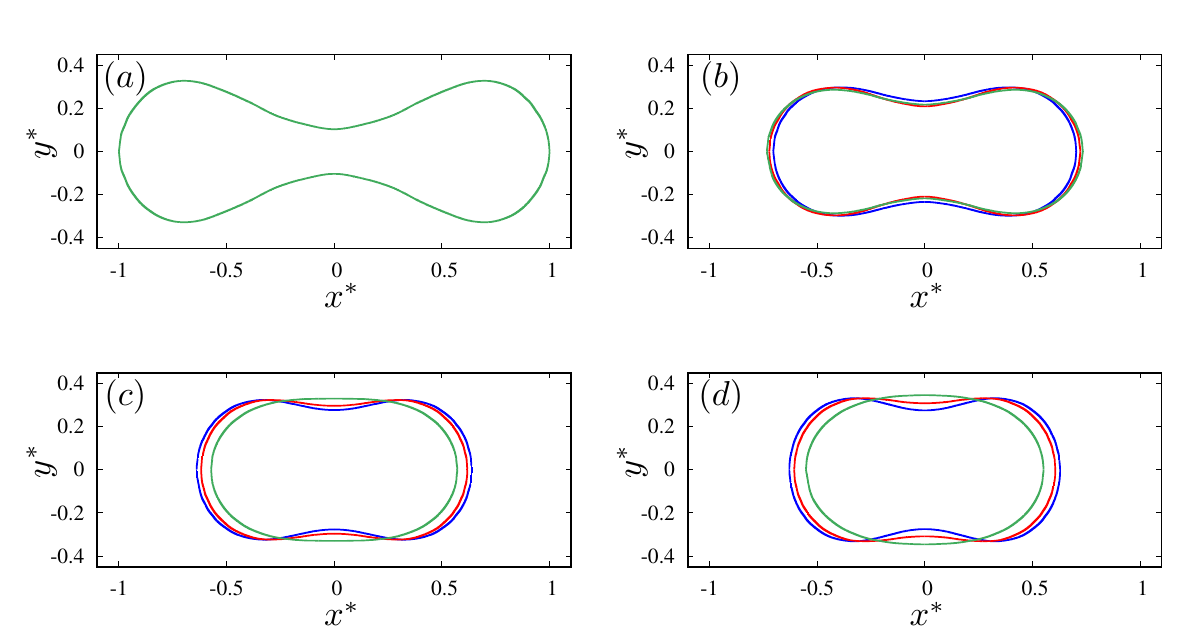}
\caption{Shape evolution in the ($x^*, y^*$) plane perpendicular to the stretching direction for the axisymmetric extensional flow (capsule = \textcolor{blue}{blue}, vesicle-capsule = \textcolor{red}{red}, capsule-capsule = \textcolor{mygreen}{green}), for $ t^* = 0$  [(\textit{a}) initial shape -- plotted in \textcolor{mygreen}{green} hereafter], $1$ (\textit{b}), $3$ (\textit{c}) and $6$ (\textit{d}).}
\label{RBC_ElongationFlow_Shape_XY}
\end{center}
\end{figure}

\begin{figure}
\begin{center}
\includegraphics[width=5.4in]{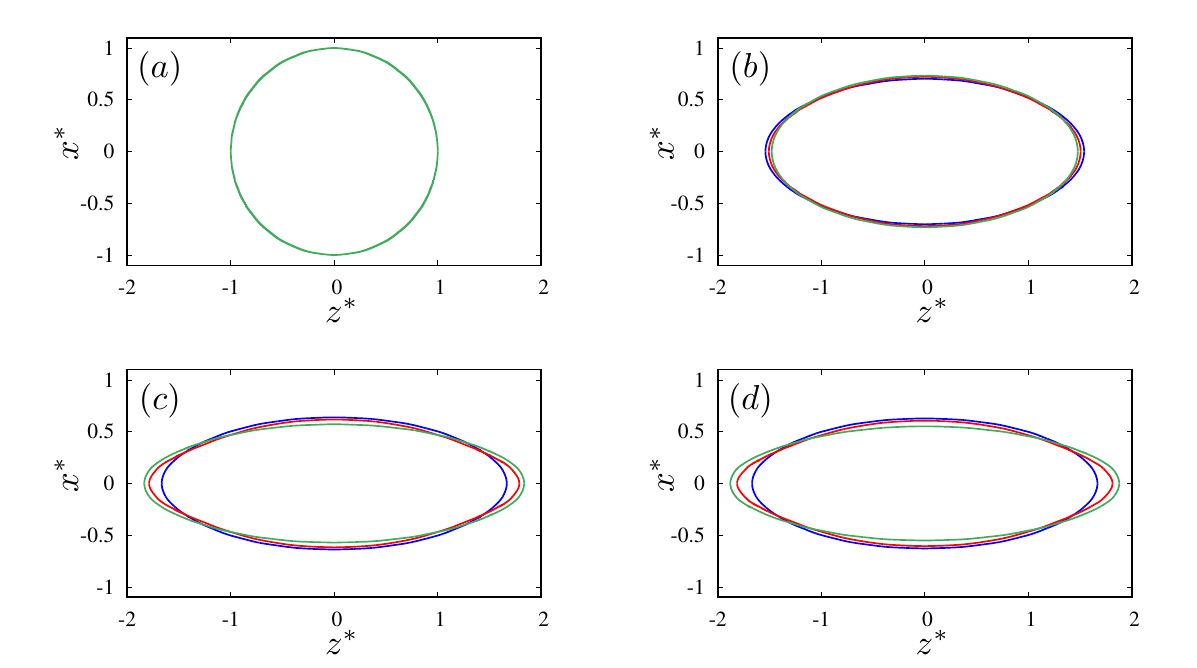}
\caption{Shape evolution in the ($x^*,z^*$) stretch plane for the axisymmetric extensional flow (capsule = \textcolor{blue}{blue}, vesicle-capsule = \textcolor{red}{red}, capsule-capsule = \textcolor{mygreen}{green}), for $ t^* = 0$  [(\textit{a}) initial shape], $1$ (\textit{b}), $3$ (\textit{c}) and $6$ (\textit{d}).}
\label{RBC_ElongationFlow_Shape_XZ}
\end{center}
\end{figure}

\begin{figure}
\begin{center}
\includegraphics[width=5.4in]{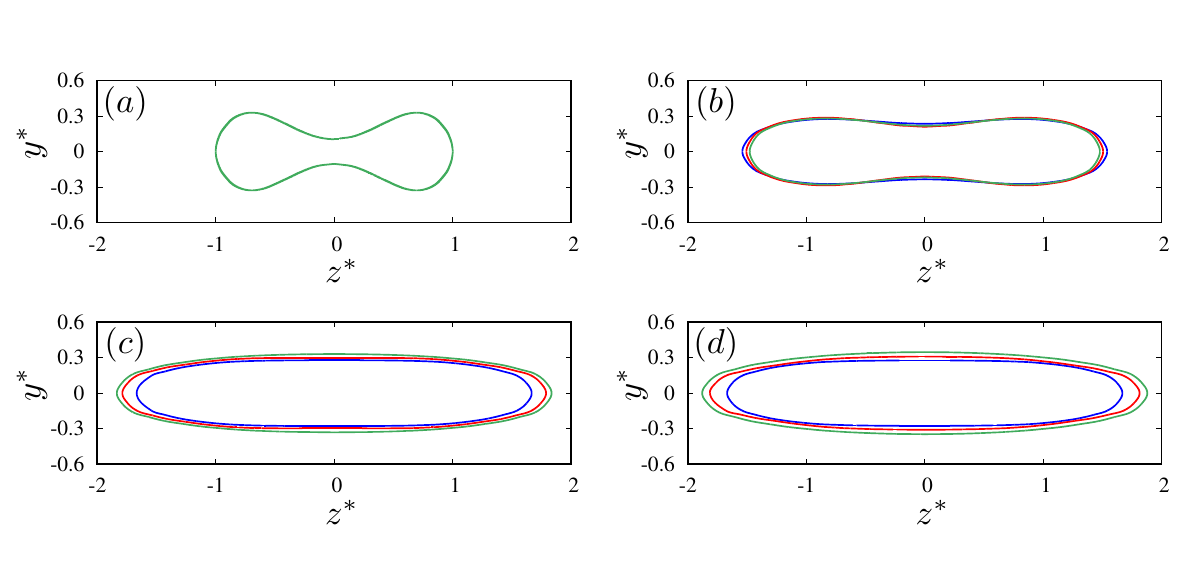}
\caption{Shape evolution in the ($y^*,z^*$) stretch plane for the axisymmetric extensional flow (capsule = \textcolor{blue}{blue}, vesicle-capsule = \textcolor{red}{red}, capsule-capsule = \textcolor{mygreen}{green}), for $ t^* = 0$  [(\textit{a}) initial shape], $1$ (\textit{b}), $3$ (\textit{c}) and $6$ (\textit{d}).}
\label{RBC_ElongationFlow_Shape_YZ}
\end{center}
\end{figure}

When sliding is allowed, the  double-layer models deviate significantly from the  single-layer capsule model, with a much higher elongation. In dimensionless time $t^*$, the capsule model reaches a state of permanent deformation at $t^*=3$, while the double-layer models continue to stretch out to $t^*=6$. Despite the slightly lower elongation of the vesicle-capsule model due to its better consideration of surface incompressibility, it remains close to that of the capsule-capsule model. The evolution of the RBC's shape is illustrated by the cross-sectional profiles in the three planes of symmetry in figures \ref{RBC_ElongationFlow_Shape_XY}, \ref{RBC_ElongationFlow_Shape_XZ} and \ref{RBC_ElongationFlow_Shape_YZ} at  $t^*=0$ (initial biconcave shape), $1$, $3$ and $6$. In all models, the dimple is reduced under stretching, with complete flattening observed in the capsule-capsule model, consistent with its higher elongation.

\begin{figure}
\begin{center}
\includegraphics[width=5.4in]{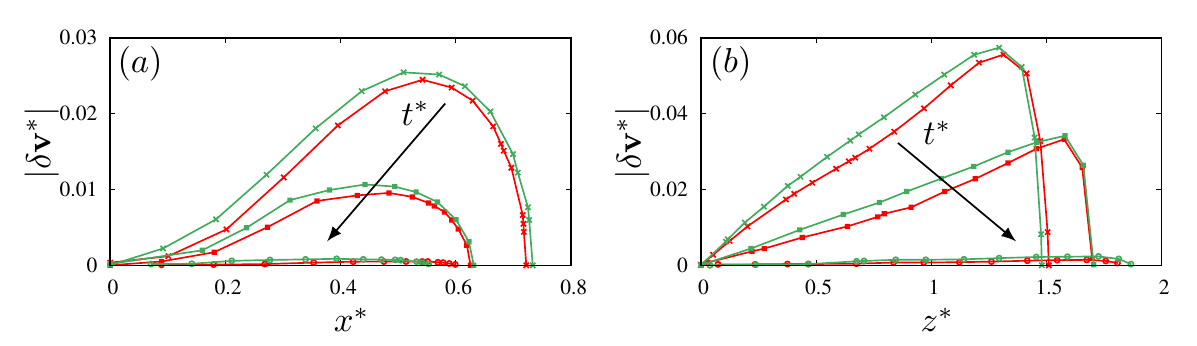}
\caption{Decrease of the sliding velocity $\delta \textbf{v}^*=\boldsymbol{v}^{*BL} - \boldsymbol{v}^{*SC}$ for $t^*=1$, 2 and $6$  for the axisymmetric extensional flow (vesicle-capsule = \textcolor{red}{red}, capsule-capsule = \textcolor{mygreen}{green}), in function of $x^*$ in the ($x^*,y^*$) plane (\textit{a}) and of $z^*$ in the ($y^*,z^*$) plane (\textit{b}).}
\label{RBC_ElongationFlow_Sliding}
\end{center}
\end{figure}

The mechanical properties of all models are identical, resulting in similar behaviour when sliding is prohibited in the double-layer strategies. However, when sliding is allowed, the observed variation in behaviour must be explained by this additional degree of freedom. To quantify the intensity of sliding, figure \ref{RBC_ElongationFlow_Sliding} displays the dimensionless velocity difference between the cytoskeleton and lipid bilayer, which shows that sliding is maximal in the initial stages and disappears completely when the stationary state is reached. The sliding velocity in the ($x^*, y^*$) plane as a function of $x^*$ (figure \ref{RBC_ElongationFlow_Sliding}\textit{a}) and in the ($y^*,z^*$) plane as a function of $z^*$ (figure \ref{RBC_ElongationFlow_Sliding}\textit{b}) is given for $t^*=1$, $2$ and $6$. The curves of sliding velocity as a function of $z^*$ in the ($x^*,z^*$) plane (not shown here) are similar to those in the ($y^*,z^*$) plane, the intensity is slightly lower, with a maximum of approximately $0.04$ instead of $0.06$, and the sliding is almost identical for both models at $t^*=1$.

\begin{figure}
\begin{center}
\includegraphics[width=5.3in]{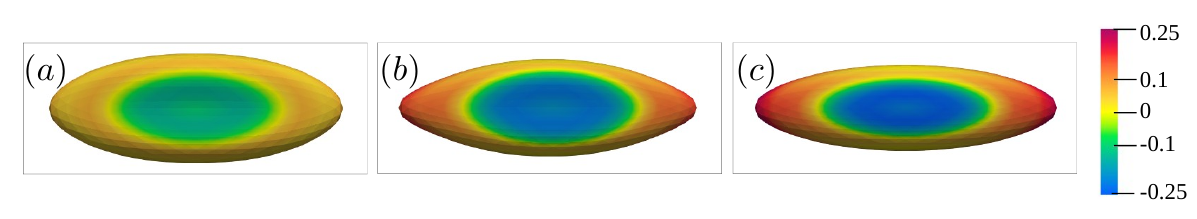}
\caption{Relative cytoskeleton area change (represented by the colour code between $-0.25$ in blue and $0.25$ in red) for the axisymmetric extensional flow at $ t^* = 6 $.  $ C^{SK} = 80 $ for the capsule model (\textit{a}) and $ C^{SK} = 2 $ for the capsule representing the cytoskeleton in the vesicle-capsule model (\textit{b}) and capsule-capsule model (\textit{c}).}
\label{RBC_ElongationFlow_Area}
\end{center}
\end{figure}

One explanation for the observed behaviour concerns the degree to which the surface incompressibility of the lipid bilayer is transmitted to the cytoskeleton. In the capsule model, the transmission is total since it is not possible to uncouple the cytoskeleton from the lipid bilayer. This is reflected by a value of the $C^{SK}$ coefficient of the Skalak model as large as possible, i.e. 80 in our simulations. On the other hand, in the double-layer models, the cytoskeleton can relax this constraint, thanks to the possibility of sliding. For these two models, $C^{SK}=2$ for the capsule which represents the cytoskeleton. Figure \ref{RBC_ElongationFlow_Area} compares the local relative area variation for the cytoskeleton between the three models. The local area variation is normalized by the total surface area of the RBC at rest, i.e. at the initial time. Although the variations are stronger and very comparable for the vesicle-capsule and capsule-capsule models, this explanation alone does not account for the observed behaviour.

\begin{figure}
\begin{center}
\includegraphics[width=3.5in]{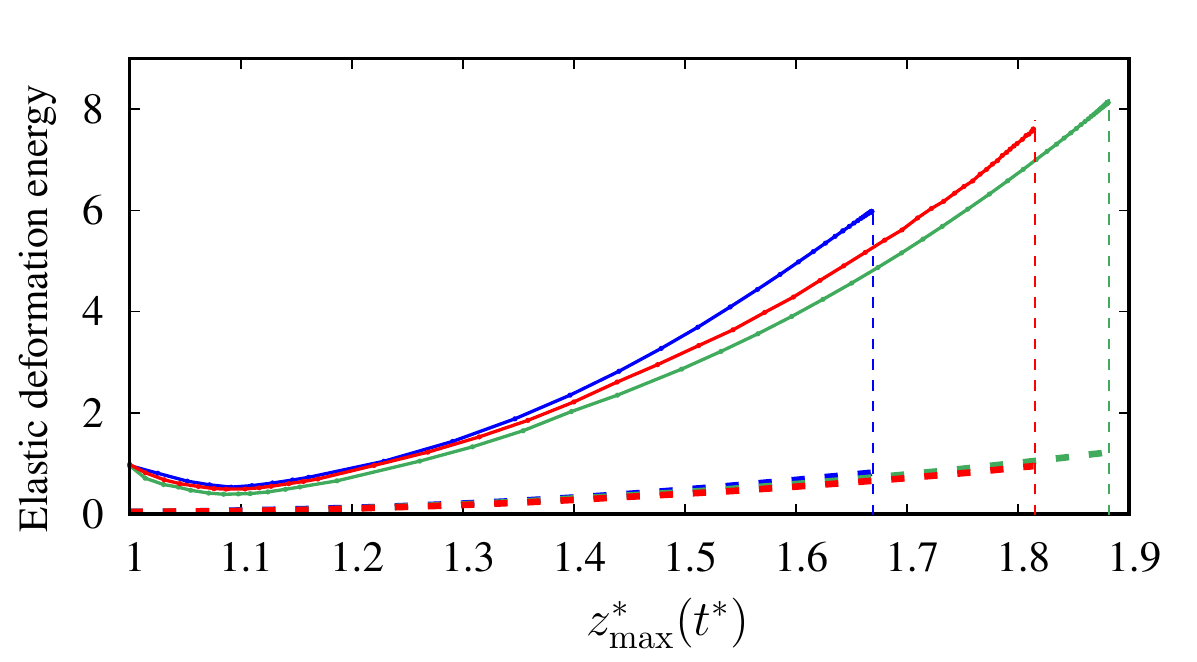} 
\caption{Elastic deformation energy ($\mathcal{W}^{BL}+\mathcal{W}^{SC}$) as a function of the RBC elongation, given by the normalised position of the RBC's tip  $z^*_\mathrm{max}$ at normalised time $t^*$ (capsule = \textcolor{blue}{blue}, vesicle-capsule = \textcolor{red}{red}, capsule-capsule = \textcolor{mygreen}{green}). Solid lines: total elastic energy (lipid bilayer + cytoskeleton). Dashed lines: bending energy contribution ($\mathcal{W}^{H}$).}
\label{RBC_ElongationFlow_Energy}
\end{center}
\end{figure}

The stress relaxation in the cytoskeleton offered by the possibility of sliding is more general, as it is not just about the local variation in the area. Instead, the cytoskeleton can better manage the whole deformation state imposed on it by the displacement of the RBC surface to which it is subject.  The possibility of sliding gives it complete freedom to optimise its elastic stress state by the tangential movement to the surface. Moreover, there is a flow amplification effect since the intensity of the velocity component according to the direction of the flow increases linearly in $|z^{*}|$. By reducing its deformation energy, the RBC can stretch further, and its two tips venture into regions where the intensity of the stretching velocity is greater, in accordance with the linear increase of the latter. 

Figure \ref{RBC_ElongationFlow_Energy} provides validation for this scenario, with a comparison of the elastic strain energy evolution for the whole RBC membrane (lipid bilayer + cytoskeleton). The graph shows the deformation energy as a function of elongation, which is characterised by the position of the RBC tip in the stretching direction, denoted as $z^*_\mathrm{max}$. This position is a nonlinear function of time, and the maximum elongation is reached when $z^*_\mathrm{max}$ stabilises.
When sliding is not allowed, the three evolution curves coincide, as seen in the blue curve obtained for the capsule model. In contrast, when sliding is allowed, the curves 
differ from each other, with the deformation energy being highest for the capsule model and lowest for the capsule-capsule model. The evolution curve for the vesicle-capsule model falls within the envelope formed by the other two curves, illustrating the correlation between the growth of deformation energy and the intensity of elongation. The final $z^*_\mathrm{max}$ value for the capsule model is lower than 1.7, while it is higher for the vesicle-capsule and capsule-capsule models, reaching almost 1.9 for the latter. The bending contribution to the deformation energy is also plotted on the graph, but its influence is found to be negligible, with the three dashed lines being nearly identical. 

It is interesting to note that the three curves of the total deformation energy all show a minimum for the same elongation value of $z^*_\mathrm{max} \approx 1.1$, which is related to the chosen reference shape for the cytoskeleton, namely quasi-spherical. The deformation energy of the cytoskeleton decreases as the RBC comes closer to its reference shape, resulting in a minimum value for the deformation energy. If the discocyte reference shape is chosen instead, the curves no longer show a minimum. However, apart from this point, the evolution of the deformation energy is similar for both reference shapes, and the conclusion drawn from them is identical. The elastic strain energy curves are also found to be very close when a spontaneous curvature is considered for the quasi-spherical reference shape. Overall, these findings provide important insights into the relationship between deformation energy and elongation for different RBC models.

\section{Comparison of RBC modelling strategies on shear flow}
\label{ShearFlow}

\begin{figure}
\begin{center}
\includegraphics[width=3.5in]{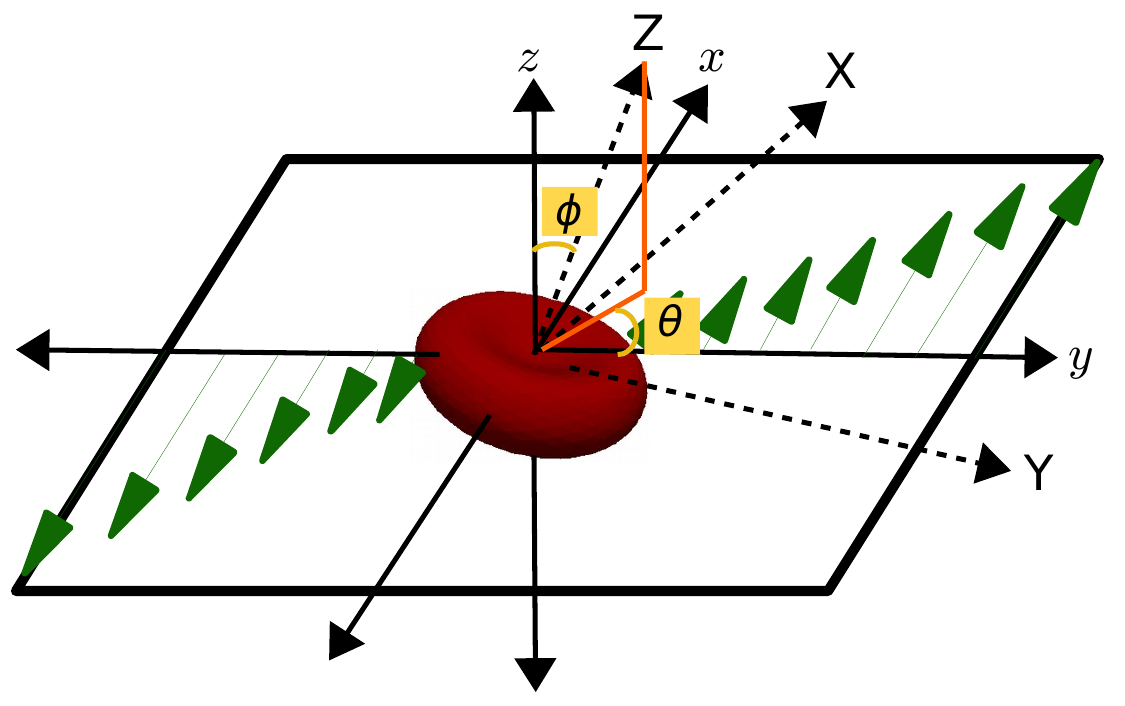}
\caption{Schematic representation of the shear flow configuration [($x$, $y$, $z$) frame] with an RBC [($X$, $Y$, $Z$) frame]. The angle $\phi$ is the orbit angle, i.e. the angle between the vorticity axis of the flow ($z$ axis) and the symmetry axis of the particle ($Z$ axis). The angle $\theta$ is the inclination angle between the direction of the velocity gradient ($y$ axis) and the projection in the shear plane of the particle's axis of symmetry ($Z$ axis).}
\label{RBC_ShearFlow}
\end{center}
\end{figure}

The simple shear flow is the most commonly used configuration for characterising the dynamics of a flowing RBC. By orienting the reference frame such that the $x$ axis corresponds to the direction of flow and the $y$ axis corresponds to that of the velocity gradient, the velocity field can be written as $ \boldsymbol{v} = v(y) \boldsymbol{e}_x $, with $ v(y) = \dot{\gamma}y $, as shown in figure \ref{RBC_ShearFlow}. The $(x, y)$ plane corresponds to the shear plane, and the $z$ axis corresponds to the vorticity axis. The intensity of the shear rate $\dot{\gamma}$ is the single operating parameter that characterises this plane flow. For a suspended particle, such as an RBC, $\dot{\gamma}$ is a hydrodynamic forcing characterised by a viscous stress $\tau = \eta_{ext}\dot{\gamma}$, where $\eta_{ext}$ is the viscosity of the suspending fluid.

Characterising the dynamics of the RBC requires considering its properties as a soft object, in addition to the characteristics of the flow. The first element determining its rigidity is the viscosity contrast $\lambda = \eta_{int}/\eta_{ext}$. Under physiological conditions, $\lambda$ is typically greater than unity because the viscosity of the cytosol of a healthy RBC can vary from $6$ to $20$ mPa\,s~\citep{mohandas_red_2008, williams_internal_2009}, whereas that of the plasma is only about $1.5$ mPa\,s [between $1$ and $1.3$ at body temperature~\citep{kesmarky_plasma_2008}]. The value of $6$ mPa\,s is generally considered to be characteristic of the cytosol viscosity of a young and healthy RBC at body temperature. All simulations were carried out with $\eta_{int} = 10$ mPa\,s, the value at room temperature. However, most experimental studies have been carried out with much more viscous outside fluids, resulting in characteristic viscosity contrasts below unity.

The second characteristic of the RBC's stiffness is related to the elastic properties of the RBC membrane, via its shear modulus $\mu_s$. Its contribution
must be compared to that of the hydrodynamic strength, which gives rise to the introduction of the capillary number $ Ca(\dot{\gamma},\mu_s) = \eta_{ext}\dot{\gamma}R/\mu_s = \tau/\tau_\mathrm{ref} $, the ratio of the viscous stress $\tau = \eta_{ext}\dot{\gamma}$ and $\tau_\mathrm{ref} = \mu_s/R$, which characterises the intensity of the elastic response of the cytoskeleton. 
Note that with our choice of time reference scale $t_\mathrm{ref} = \eta_{ext}R/\mu_s$, the capillary number may also be defined as $Ca(\dot{\gamma},\mu_s) = \dot{\gamma}t_\mathrm{ref}$.

The dynamic regimes of RBC, and more generally of a capsule, in shear flow can be represented in the plane $ [\lambda, Ca(\dot{\gamma},\mu_s)]$. For a vesicle, the reference elastic property is the bending modulus $k_b$. The capillary number used therefore is $ Ca(\dot{\gamma},k_b) = \eta_{ext}\dot{\gamma}R^3/k_b $, which is related to the capillary number for capsules by the dimensionless Von Karman number $\mathcal{K} = k_b/R^2\mu_s = (k_b/R^3)/{\tau_\mathrm{ref}}$, reflecting the relative importance of curvature elasticity (out-of-plane) versus shear elasticity (in-plane). For RBCs, the Von Karman number is approximately $5\times10^{-3}$ based on average values from table 1 of \cite{levant_intermediate_2016}. 

We use normalised quantities (indicated by a star) and dimensionless input data in our simulations. All surface density of force quantities is normalised by the reference elastic stress $\tau_\mathrm{ref}$. Hence, the actual values of $\mu_s$ and $k_b$ are not provided. Instead, the former is specified using the capillary number $Ca(\dot{\gamma},\mu_s)$, while the latter is specified using the Von Karmann number $\mathcal{K}$. 

Successive experimental studies have continued to enrich the RBC's phase diagram, and a synthesis of the experimental observations is proposed in \citet{minetti_dynamics_2019}. Studies often assume the axis of symmetry of the RBC remains in the shear plane. Under these conditions, the RBC's dynamics can be characterised by the inclination angle $\theta$ (see figure~\ref{RBC_ShearFlow}) and the Taylor deformation parameter $D= (L - B)/(L+B)$, in which $L$ and $B$ are the major and minor axes of the ellipsoid having the same moment of inertia as the RBC. At low values of $\tau$, the RBC exhibits tumbling dynamics. As $\tau$ increases, the axis of symmetry moves closer to the direction of the velocity and the deformation becomes more intense, resulting in the transition from tumbling to tank-treading dynamics. The inclination angle and the Taylor parameter take constant values in the tank-treading regime, with periodic variations around their mean values (swinging) but with decreasing amplitude at higher $\tau$. However, experimental studies indicate that an increase in the shear rate is more likely to cause the axis of symmetry to drift out of the shear plane, leading to rolling dynamics and the appearance of new dynamic regimes showing RBCs with stomatocyte and then multilobe shapes \citep{mauer_flow-induced_2018}. The critical transition capillary number curve $Ca_c(\dot{\gamma},\mu_s)$ reaches its minimum when $\eta_{ext}$ is greater than 20 mPa s \citep{fischer_threshold_2013}, making it easier to reach the tank-treading regime at high values of $\eta_{ext}$. As we move towards physiological values of $\eta_{ext}$, i.e. at $ \lambda > 1$, the curve $Ca_c(\dot{\gamma},\mu_s)$ sharply increases and the transition can only be reached by imposing very high shear rates.

To simplify our analysis, we have assumed that the axis of symmetry of the RBC remains in the shear plane. Although this assumption may not hold for all transition routes, it is sufficient for our purpose, which is to compare modelling strategies at representative points of RBC dynamics. Moreover, as demonstrated in \cite{levant_intermediate_2016}, the relationship between simple shear flow and plane linear flow allows for this simplified configuration. Another reason to restrict to this simpler configuration is that the characteristic drift time of the RBC symmetry axis out of the shear plane can be large relative to the time scale of RBC dynamics for large $\eta_{ext}$ and small $\dot{\gamma}$ \citep{levant_intermediate_2016}. In addition, under physiological conditions, stable orbits widen as the shear rate decreases, and 
all orbits become stable below a certain value of $\tau$, which is approximately $10^{-2}$ Pa [as shown in figure 14\textit{c} in \cite{minetti_dynamics_2019}]. This trend is consistent at both low and high $\eta_{ext}$ values, as long as we consider equivalent viscous stress. Therefore, exploring the low $\tau$ region is easier when $\eta_{ext}$ is small, as in physiological conditions.

To comprehensively explore the diverse dynamic regimes of an RBC in shear flow, our comparative study focuses on four key points $ [\lambda,    Ca(\dot{\gamma},\mu_s)] $ of the phase diagram. The first point corresponds to the tumbling regime, which is a stable regime for an RBC with its axis of symmetry in the flow plane under physiological conditions. The second point corresponds to the tank-treading regime, which is observed at $\eta_{ext}$ values greater than $20$ mPa s. We consider both possible orientations of the axis of symmetry of the RBC, aligned with the axis of vorticity or in the shear plane. The third point is located in the intermittency region, where both tumbling and tank-treading dynamics coexist. Finally, the last point is selected under physiological conditions but with a high shear rate to compare the modelling strategies in the regime of very high deformations. 

\subsection{Effect of modelling strategy on tumbling dynamics}
\label{sec:pt1}

\begin{figure}
\begin{center}
\includegraphics[width=5.5in]{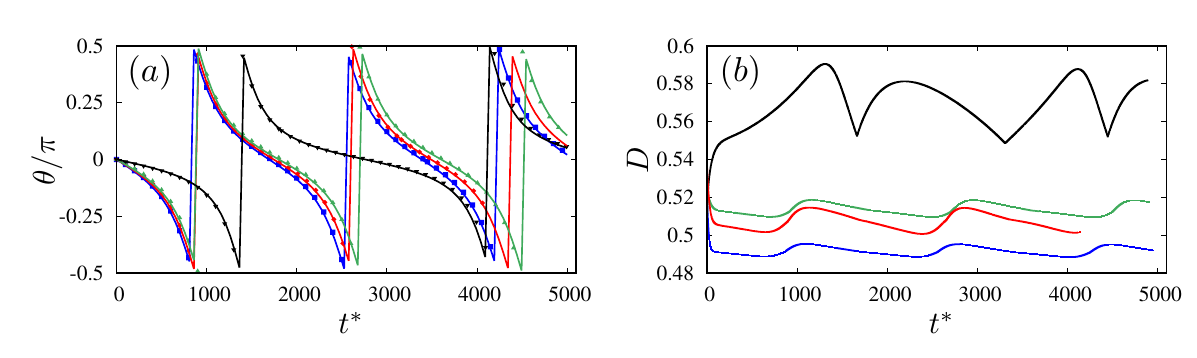}
\caption{(\textit{a}) Time evolution of the inclination angles $\theta/\upi$ on tumbling dynamics at $[ \lambda = 6.67, Ca(\dot{\gamma},\mu_s) = 5\times 10^{-3} ]$. Points: simulations (vesicle = \textcolor{black}{black}, capsule = \textcolor{blue}{blue}, vesicle-capsule = \textcolor{red}{red}, capsule-capsule = \textcolor{mygreen}{green}). Solid lines: Jeffery's theory (second equation of \ref{Jeffery}) for fitted $r$ on frequency criteria ($ r = 2.25 $ for capsule, $ r = 2.35 $ for vesicle-capsule, $ r = 2.46 $ for capsule-capsule, and $ r = 4.15 $ for vesicle). (\textit{b}) Time evolution of the deformation parameter $D$.}
\label{TU}
\end{center}
\end{figure}

Our first investigation point considered a viscous stress of 0.01 Pa and an external viscosity of 1.5 mPa s, corresponding to the characteristic dimensionless numbers $\lambda = 6.67$ and $Ca(\dot{\gamma},\mu_s) = 5 \times 10^{-3}$. We found that the inclination angle of the RBC (figure \ref{TU}\textit{a}) undergoes a tumbling motion, with the spinning frequency around the vorticity axis varying based on the model used. Interestingly, the capsule model exhibited the highest frequency, while the vesicle model's rotation frequency was almost half that of the other models.

Additionally, the Taylor deformation parameter $D$ (figure \ref{TU}\textit{b}) highlighted a direct relationship between an object's stiffness and its rotation frequency, with the effect being particularly amplified in the vesicle model, which lacks shear elasticity. Further analysis showed that the stiffening effect of the surface incompressibility constraint was clearly demonstrated in the comparison between the capsule-capsule and vesicle-capsule models. 

Since the RBC's deformation is small, this dynamic regime is close to that of a rigid particle, and Jeffery's theory~\citep{jeffery_motion_1922} 

\begin{equation}
\label{Jeffery}
r\tan\phi = \frac{C^{orbit}}{\sqrt{r^{-2}\cos^2\theta+\sin^2\theta}} \, , \quad r\tan\theta = \tan \frac{\dot{\gamma}t}{r+r^{-1}}
\end{equation}
holds well, where $C^{orbit} = r\tan\phi_0$ is the orbit parameter and $r$ is the particle's aspect ratio. 
The prolate and oblate shapes are characterised by $r < 1$ and $r > 1$, respectively, with $r$ appearing in the equation $X^2 + Y^2 + r^2Z^2 = 1$ for the surface of an ellipsoidal object. It's important to note that, in this theory, all orbits are probable and determined solely by the initial value $\phi_0$ of the angle $\phi$. To verify Jeffery's theory, we superimposed the curves obtained for each modelling strategy using the second equation of \ref{Jeffery} as solid lines (figure \ref{TU}\textit{a}), with the aspect ratio $r$ adjusted to reproduce the oscillation frequency of the model.

\subsection{Effect of modelling strategy on tank-treading dynamics}
\label{sec:pt2}

\begin{figure}
\begin{center}
\includegraphics[width=5.3in]{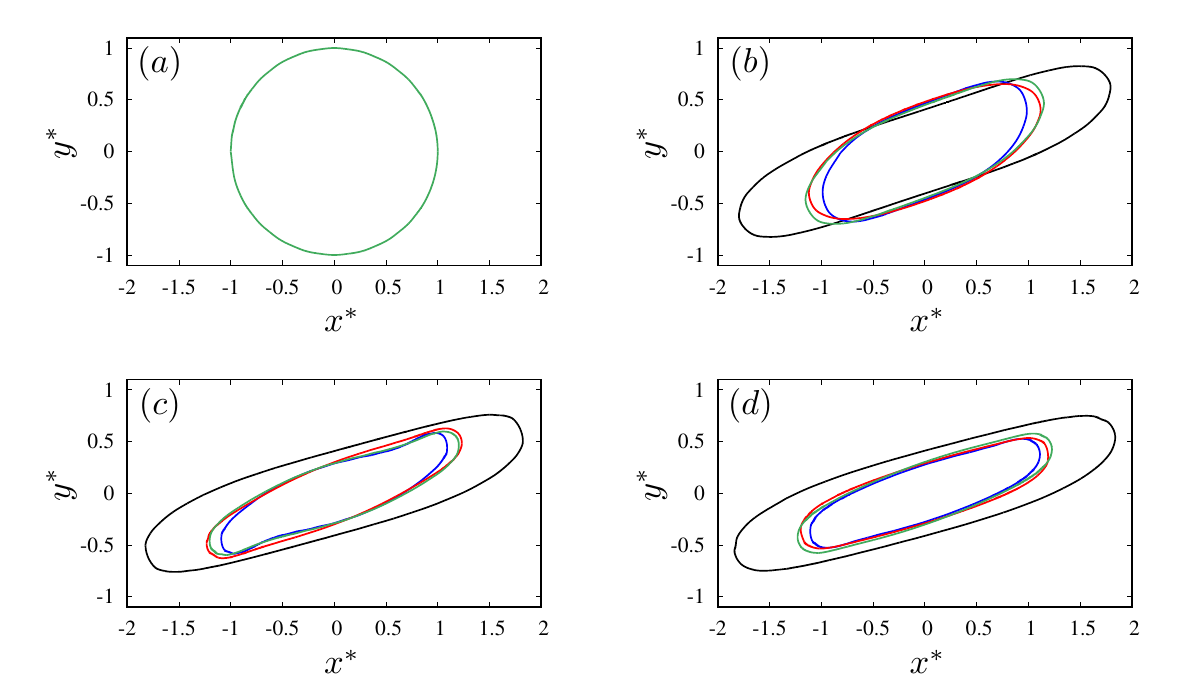}
\caption{
Shape evolutions [$(x^*,y^*)$ plane sectional drawing] on tank-treading dynamics at $[ \lambda = 0.4, Ca(\dot{\gamma},\mu_s) = 0.5 ]$ in the transient phase at $ t^* = 0 $ (\textit{a}), $ t^* = 6 $ (\textit{b}), $ t^* = 11 $ (\textit{c}) and $ t^* = 15 $ (\textit{d}), for the case when the RBC's symmetry axis is initially aligned with the axis of vorticity (vesicle = \textcolor{black}{black}, capsule = \textcolor{blue}{blue}, vesicle-capsule = \textcolor{red}{red}, capsule-capsule = \textcolor{mygreen}{green}). }
\label{TT_Shape_Transient_Z_XY}
\end{center}
\end{figure}

\begin{figure}
\begin{center}
 \includegraphics[width=4.0in]{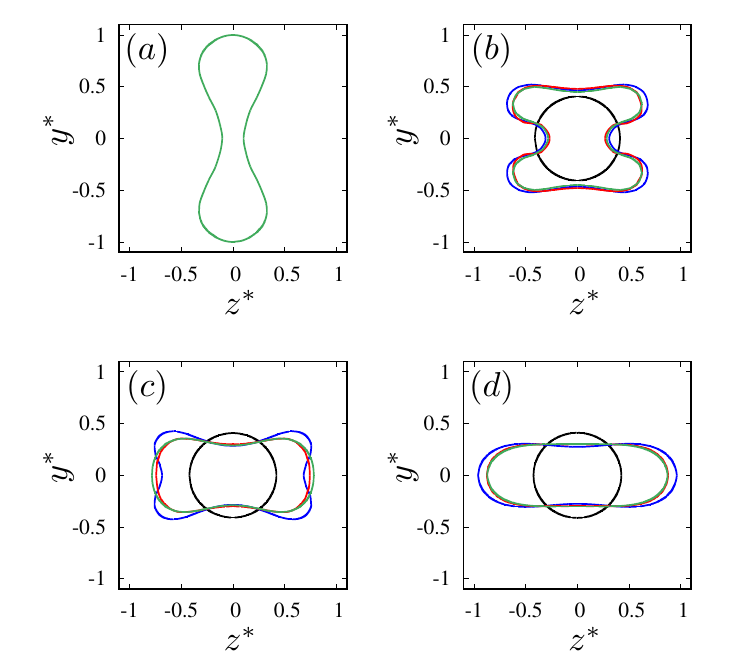}
\caption{
Shape evolutions [$(y^*,z^*)$ plane sectional drawing] on tank-treading dynamics at $[ \lambda = 0.4, Ca(\dot{\gamma},\mu_s) = 0.5 ]$ in the transient phase at $ t^* = 0 $ (\textit{a}), $ t^* = 6 $ (\textit{b}), $ t^* = 11 $ (\textit{c}) and $ t^* = 15 $ (\textit{d}), for the case when the RBC's symmetry axis is initially aligned with the axis of vorticity (vesicle = \textcolor{black}{black}, capsule = \textcolor{blue}{blue}, vesicle-capsule = \textcolor{red}{red}, capsule-capsule = \textcolor{mygreen}{green}).}
\label{TT_Shape_Transient_Z_YZ}
\end{center}
\end{figure}

\begin{figure}

\begin{center}
\includegraphics[width=5.3in]{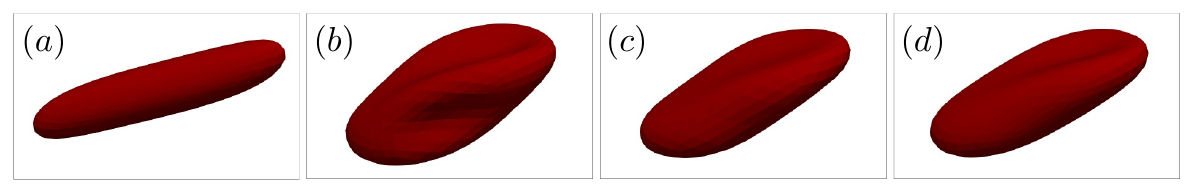}
\caption{
Shapes (3D view) on tank-treading dynamics at $[ \lambda = 0.4, Ca(\dot{\gamma},\mu_s) = 0.5 ]$
in the transient phase at $ t^* = 11 $, for the case when the RBC's symmetry axis is initially aligned with the axis of vorticity.
(\textit{a}) vesicle, (\textit{b}) capsule, (\textit{c}) vesicle-capsule, (\textit{d}) capsule-capsule.}
\label{TT_Shape_Transient_Z_3D}
\end{center}
\end{figure}

In the second point, we investigated the effects of increased external viscosity ($\eta_{ext} = 25$ mPa s) and high shear rate ($\tau = 1.06$ Pa) on RBCs, characterised by the dimensionless numbers $ \lambda = 0.4 $ and $ Ca(\dot{\gamma},\mu_s) = 0.5 $. When the RBC axis of symmetry is initially aligned with the axis of vorticity, the transient phase during which it adapts its shape is longer. Figures \ref{TT_Shape_Transient_Z_XY} and \ref{TT_Shape_Transient_Z_YZ} show the shape evolution during this phase in the $(x^*,y^*)$ and $(z^*,y^*)$ planes, respectively. All models follow a similar transition pattern with equivalent times, except for the vesicle model which undergoes a large deformation and then lengthens. However, a difference in behaviour can be observed between single-layer and double-layer strategies. While the capsule model still shows a curvature inversion at $ y^* = 0 $ in the $(y^*, z^*)$ plane at $ t^* = 11 $, the double-layer models do not. Figure \ref{TT_Shape_Transient_Z_3D} provides a 3D view at that time.


\begin{figure}
\begin{center}
\includegraphics[width=5.4in]{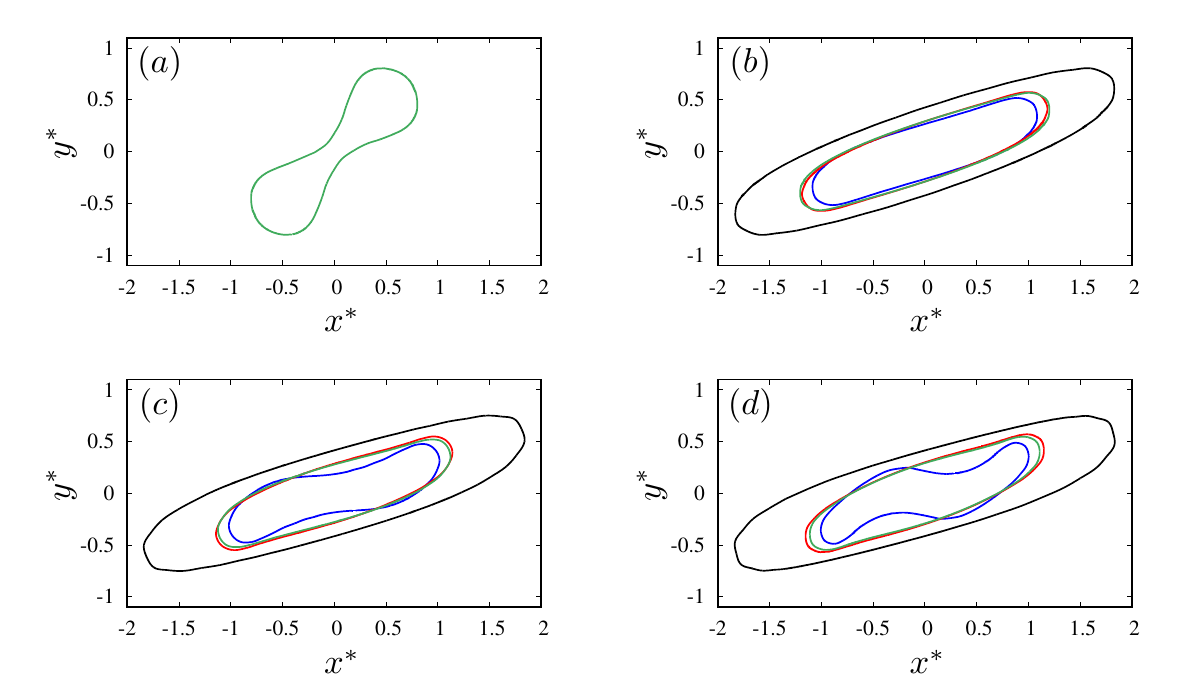}
\caption{
Shape evolutions [$(x^*,y^*)$ plane sectional drawing] on tank-treading dynamics at $[ \lambda = 0.4, Ca(\dot{\gamma},\mu_s) = 0.5 ]$ in the transition phase at $ t^* = 0 $ (\textit{a}), $ t^* = 6 $ (\textit{b}), $ t^* = 11 $ (\textit{c}) and $ t^* = 15 $ (\textit{d}), when the RBC's symmetry axis is in the shear plane at the start (vesicle = \textcolor{black}{black}, capsule = \textcolor{blue}{blue}, vesicle-capsule = \textcolor{red}{red}, capsule-capsule = \textcolor{mygreen}{green}).}
\label{TT_Shape_Transient_XY_XY}
\end{center}
\end{figure}

\begin{figure}
\begin{center}
\includegraphics[width=5.4in]{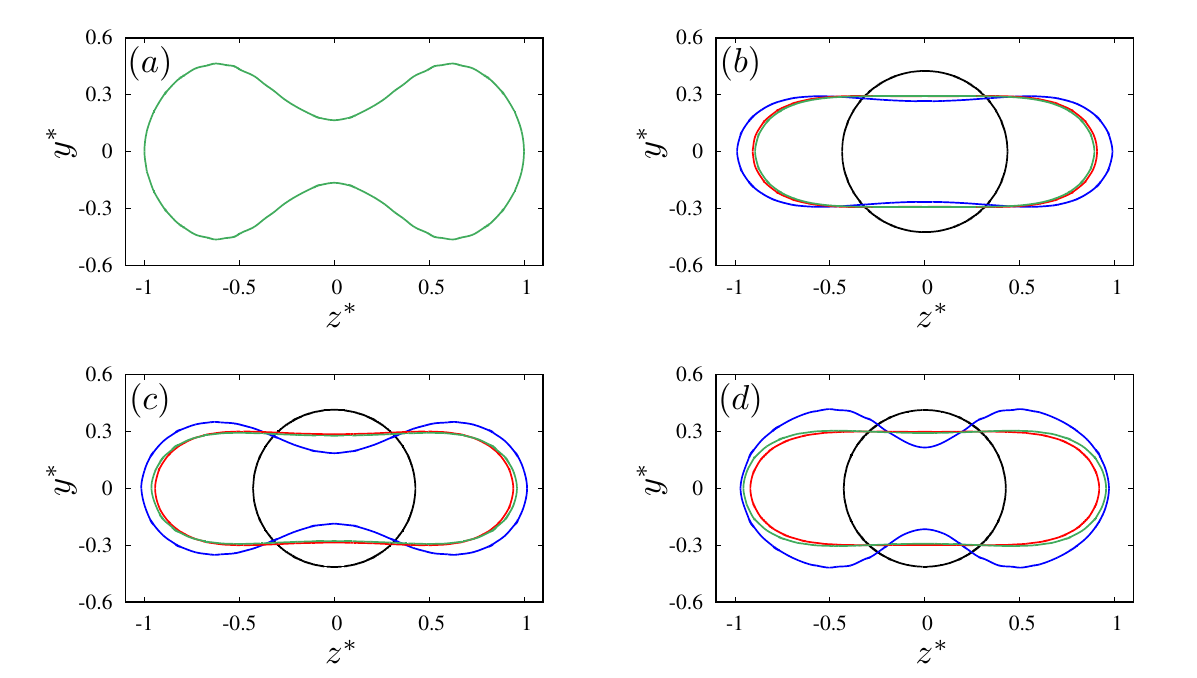}
\caption{Shape evolutions [$(y^*,z^*)$ plane sectional drawing] on tank-treading dynamics at $[ \lambda = 0.4, Ca(\dot{\gamma},\mu_s) = 0.5 ]$ in the transition phase at $ t^* = 0 $ (\textit{a}), $ t^* = 6 $ (\textit{b}), $ t^* = 11 $ (\textit{c}) and $ t^* = 15 $ (\textit{d}), when the RBC's symmetry axis is in the shear plane at the start (vesicle = \textcolor{black}{black}, capsule = \textcolor{blue}{blue}, vesicle-capsule = \textcolor{red}{red}, capsule-capsule = \textcolor{mygreen}{green}).}
\label{TT_Shape_Transient_XY_YZ}
\end{center}
\end{figure}

\begin{figure}
\begin{center}
\includegraphics[width=5.3in]{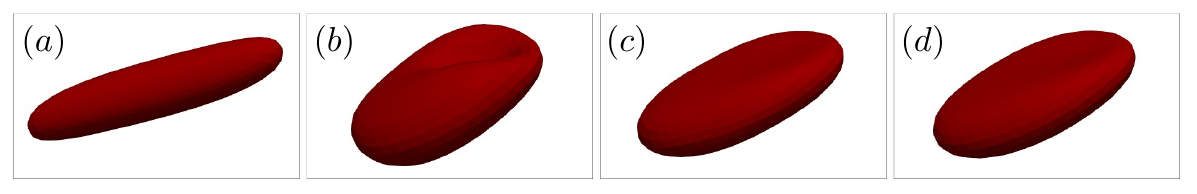}
\caption{Shapes (3D view) on tank-treading dynamics at $[ \lambda = 0.4, Ca(\dot{\gamma},\mu_s) = 0.5 ]$
in the transition phase at $ t^* = 15 $, when the RBC's symmetry axis is in the shear plane at the start
(\textit{a}) vesicle, (\textit{b}) capsule, (\textit{c}) vesicle-capsule, (\textit{d}) capsule-capsule.}
\label{TT_Shape_Transient_XY_3D}
\end{center}
\end{figure}

When the RBC's axis of symmetry is initially in the shear plane, it is already well-oriented relative to its final steady state. The transient phase of large deformations is, correspondingly, reduced. The shape evolution during this phase in the $(x^*,y^*)$ and $(z^*,y^*)$ planes is shown in figures \ref{TT_Shape_Transient_XY_XY} and \ref{TT_Shape_Transient_XY_YZ}, respectively. Figure \ref{TT_Shape_Transient_XY_3D} at $ t^* = 15 $ provides a 3D view.

\begin{figure}
\begin{center}
\includegraphics[width=5.3in]{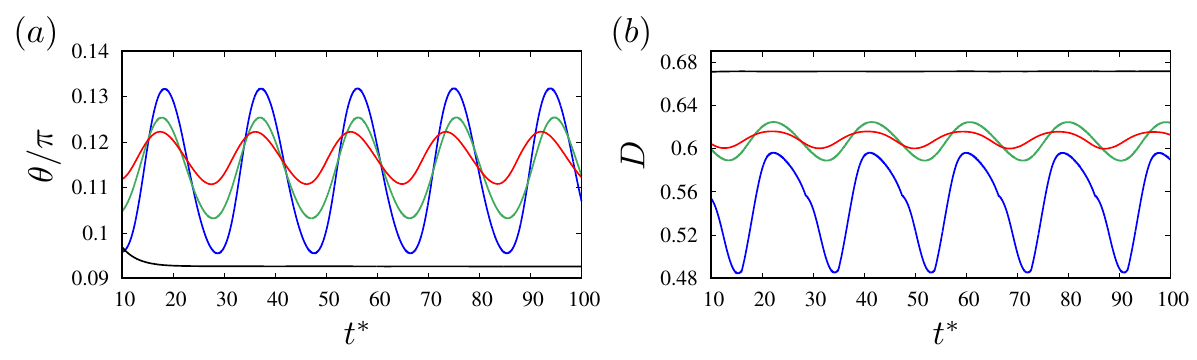}\caption{
Time evolution of the inclination angle $\theta/\upi$ (\textit{a}) and the deformation parameter $D$ (\textit{b}) on tank-treading dynamics at $[ \lambda = 0.4, Ca(\dot{\gamma},\mu_s) = 0.5 ]$,
when the RBC's symmetry axis is initially in the shear plane (vesicle = \textcolor{black}{black}, capsule = \textcolor{blue}{blue}, vesicle-capsule = \textcolor{red}{red}, capsule-capsule = \textcolor{mygreen}{green}).}
\label{TT_XY}
\end{center}
\end{figure}

After the transient phase, the dynamics are identical, regardless of the initial orientation of the symmetry axis. The evolution curves in figure \ref{TT_XY} are indistinguishable from those obtained when the axis of symmetry is aligned with the axis of vorticity, with a normalised time shift of $ \triangle t^* =11.6 $. Thus, generalisation to any initial orientation is highly likely.

Once the steady state is reached, the inclination angle's evolution (figure \ref{TT_XY}\textit{a}) reveals that the swinging is most pronounced for the capsule model and least pronounced for the vesicle-capsule model. The capsule-capsule model is in between. As expected, the vesicle model exhibits pure tank-treading motion, characterised by a constant value of the inclination angle.

The evolution of deformation via the Taylor parameter $D$ (figure \ref{TT_XY}\textit{b}) remains the most effective way to distinguish between the different models. As expected, the vesicle model has the most intense deformation and is completely separate from the other models. The double-layer models produce mean deformations of the same order, with an almost sinusoidal regularity of the Taylor parameter evolution. However, the stiffness provided by the surface incompressibility is reflected in an oscillation amplitude half as large for the vesicle-capsule model as for the capsule-capsule model. Although the mean deformation intensity is lower for the capsule model, its oscillation amplitude is two to three times greater than that of the capsule-capsule model, with the oscillations appearing to be much less symmetrical. We also note that there is an apparent increase in the rotation frequency when the lipid bilayer is modelled as an incompressible fluid film rather than a solid shell. However, it is difficult to conclude whether this increase is related to the consideration of the fluid nature of the lipid bilayer or to the more rigorous treatment of the surface incompressibility constraint.

\begin{figure}

\hspace{0.2cm} \includegraphics[width=6.5in]{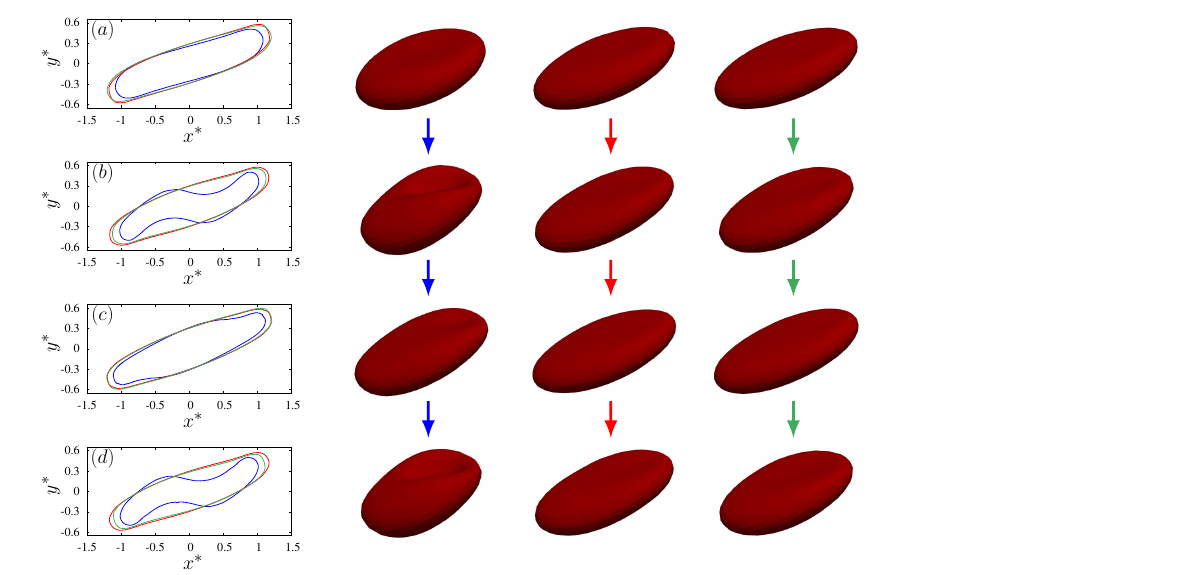}
\caption{Shape evolution on established tank-treading dynamics at $[  \lambda = 0.4, Ca(\dot{\gamma},\mu_s) = 0.5 ]$,  when the RBC's symmetry axis is initially in the shear plane. Left panel: $(x^*,y^*)$ plane sectional drawing at $ t^* = 7 $ (\textit{a}), $ t^* = 15 $ (\textit{b}), $ t^* = 20 $ (\textit{c}) and $ t^* = 33 $ (\textit{d}) (capsule = \textcolor{blue}{blue}, vesicle-capsule = \textcolor{red}{red}, capsule-capsule = \textcolor{mygreen}{green}). Right panel: corresponding 3D views for capsule (left), vesicle-capsule (middle) and capsule-capsule (right).}
\label{TT_Shape_XY}

\end{figure}

Figure \ref{TT_Shape_XY} compares the shape evolutions in the established regime, with the vesicle model excluded due to its much larger deformation. The double-layer models produce very similar shapes that are relatively stable, while the capsule model undergoes strong shape variations. In cross-section in the shear plane, the evolution of the capsule model periodically changes from an elliptical to an S-shape (breathing phenomenon).

A legitimate question is whether the transitional phase in the case where the axis of symmetry is aligned with the axis of vorticity is only a shape adaptation or whether it is accompanied by a general pivoting of the cytoskeleton. If the answer is negative, the final state reached cannot be considered completely identical to the one reached starting from the other orientation, at least from an energetic point of view. To answer this question, we tracked the $y^*$ coordinate of two markers, one initially located on the dimple in the centre of one of the two faces of the RBC and the other on the periphery. The answer is clearly negative: when the axis of symmetry is initially in the shear plane, the dimple undergoes the tank-treading movement, whereas when the axis of symmetry is initially aligned with the axis of vorticity, the marker at the periphery performs the rotation.

\subsection{Effect of modelling strategy on the transition to tank treading}
\label{sec:pt3}

Our third point of investigation lies at the upper limit of the intermittency region. For an external viscosity $\eta_{ext}=24$ mPa s, \citet{fischer_threshold_2013} identified a critical shear rate of 10 s$^{-1}$ for the transition, corresponding to a critical viscous stress of 0.24 Pa. The corresponding dimensionless numbers are $\lambda = 0.417$ and $Ca(\dot{\gamma},\mu_s) = 0.113$.

\begin{figure}
\begin{center}
\includegraphics[width=5.3in]{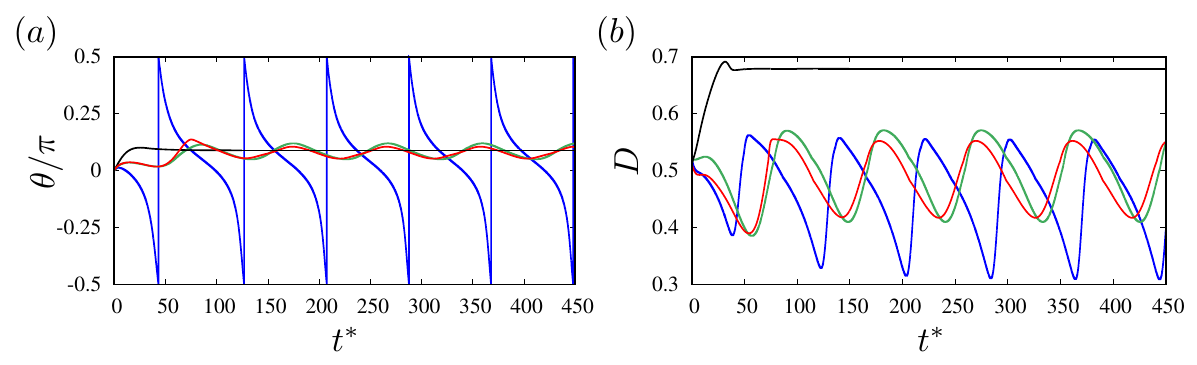}
\caption{Time evolution of the inclination angle $\theta/\upi$ (\textit{a}) and the deformation parameter $D$ (\textit{b}) in the tumbling to tank-treading transition region at $[ \lambda = 0.417, Ca(\dot{\gamma},\mu_s) = 0.113 ]$ (vesicle = \textcolor{black}{black}, capsule = \textcolor{blue}{blue}, vesicle-capsule = \textcolor{red}{red}, capsule-capsule = \textcolor{mygreen}{green}).}
\label{INT}
\end{center}
\end{figure}

Since this point is located in the transition zone, it is expected that the models' behaviour is particularly sensitive to the modelling strategy. This sensitivity is clearly demonstrated in figure \ref{INT}. The vesicle model exhibits pure tank treading with a constant inclination angle. The capsule model, on the other hand, stays in the tumbling regime. The double-layer models initially exhibit tumbling, but after one or two periods, they adopt a permanent tank-treading dynamic. Their inclination angle oscillates around a mean value similar to that of the vesicle model, with a slightly larger amplitude for the capsule-capsule model.

\subsection{Effect of modelling strategy in very high deformation regime}
\label{sec:pt4}

\citet{mauer_flow-induced_2018} reported a wide range of RBC shapes and dynamics observed in their microfluidic experiments and through simulations using two distinct techniques. Notably, the multilobe regime represents conditions closely resembling the physiological environment, characterised by an external viscosity of $\eta_{ext} = 1.5$ mPa s, at very high shear rates with $\tau = 3.19$ Pa. This corresponds to dimensionless numbers $\lambda = 6.67$ and $Ca(\dot{\gamma},\mu_s) = 1.5$. We use these simulation parameters to align with the phase diagram presented by~\citet{mauer_flow-induced_2018}, pinpointing a specific point within this multilobe regime. As anticipated, our simulations effectively yielded multilobe shapes, indicating a strong alignment between our modelling approaches and experimental observations. It's worth noting that multilobe shapes have also been observed in a study using a 2D vesicle model \citep{Abbasi_2022}, suggesting that the cytoskeleton may not be necessary for the multilobe manifestation.

\begin{figure}
\begin{center}
\includegraphics[width=5.5in]{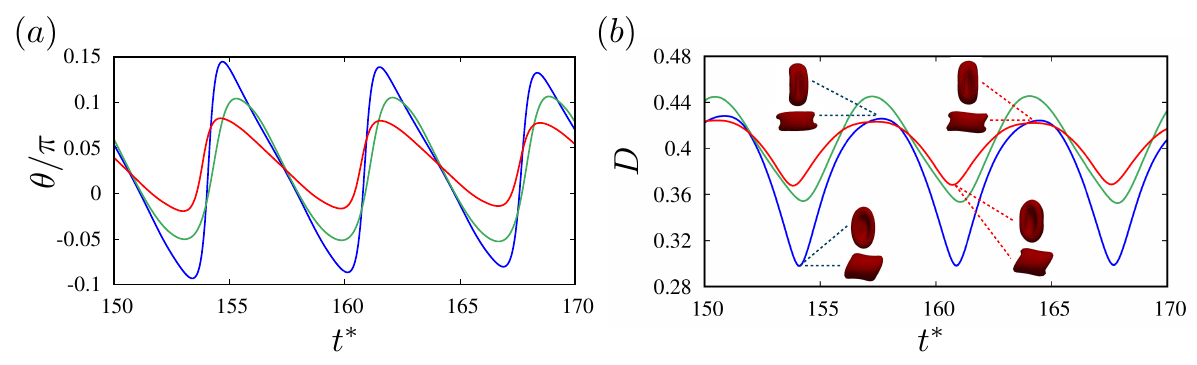}
\caption{
Time evolution of the inclination angle $\theta/\upi$ (\textit{a}) and the deformation parameter $D$ (\textit{b}) in the multilobe region at
$[ \lambda = 6.67, Ca(\dot{\gamma},\mu_s) = 1.5 ]$ 
(capsule = \textcolor{blue}{blue}, vesicle-capsule = \textcolor{red}{red}, capsule-capsule = \textcolor{mygreen}{green}).
Inserted in (\textit{b}) are $x$ and $y$ shape views for the capsule and vesicle-capsule models when $D$ is minimal and maximal.
}
\label{Multilobe}
\end{center}
\end{figure}

The evolution of the inclination angle and the Taylor deformation parameter is shown in figure \ref{Multilobe}. The deformation parameter varies between a minimum $D_\mathrm{min}$ and a maximum $D_\mathrm{max}$, resulting in different shapes depending on the modelling strategy. 

\subsection{Comparison with experimental data}
\label{sec:comp}

In \S\ref{sec:pt1}--\S\ref{sec:pt4}, we intentionally selected four points within the phase diagram of an RBC in shear flow. These points were chosen to comprehensively cover the spectrum of RBC dynamics. Across each of these chosen points, our simulations consistently replicated the anticipated shapes and dynamic responses, including the characteristic multilobe shape. These consistent agreements between our simulation results and experimental observations serve as compelling validation, affirming the effectiveness and reliability of our modelling approaches.

In this subsection, we directly compare our numerical results with existing experimental data. Specifically, we analyse the relationship between the tank-treading frequency ($f$) and the shear rate ($\dot{\gamma}$). \citet{Fischer_2007} previously reported that the frequency scales with shear rate as $f\sim \dot{\gamma}^{\beta}$, with the scaling exponent $\beta$ in the range of 0.85 to 0.95.

\begin{figure}
\begin{center}
\includegraphics[width=3.5in]{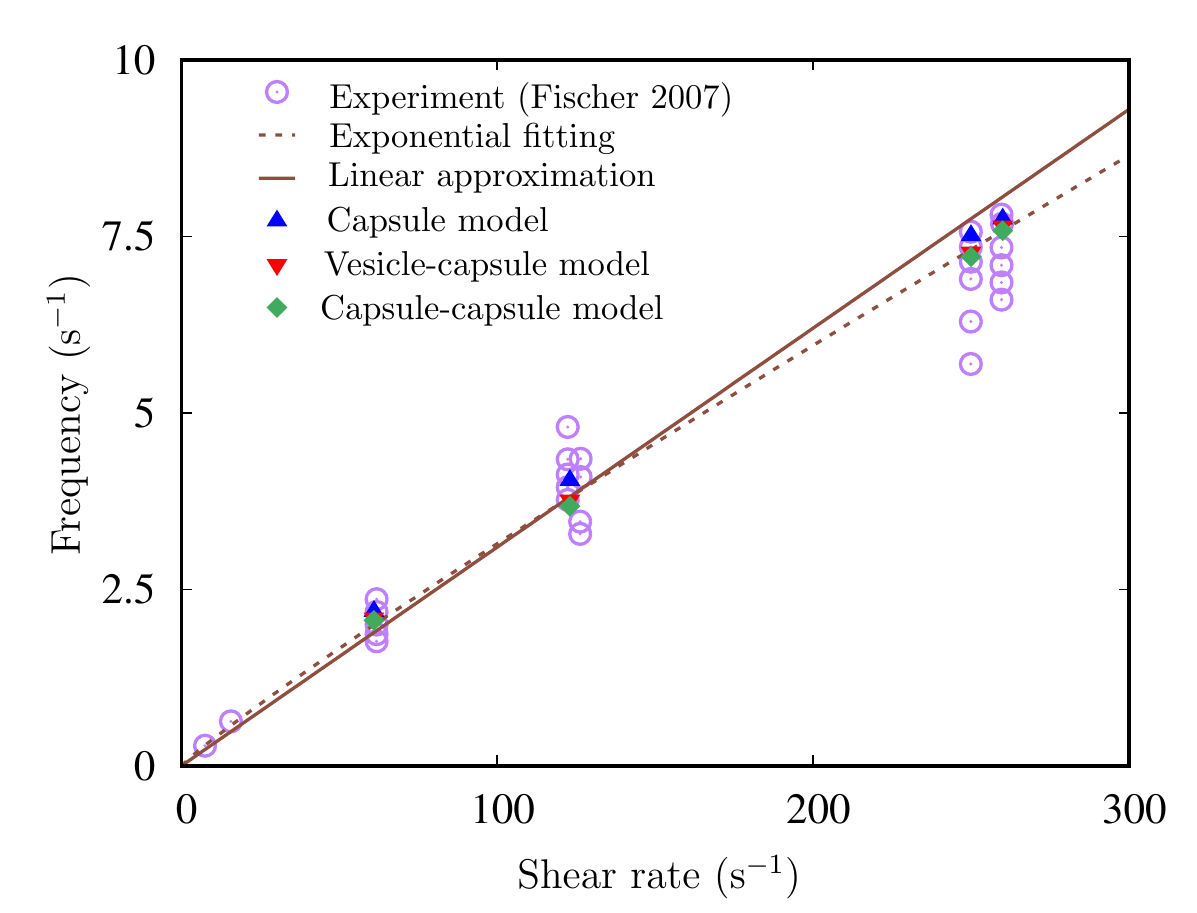}
\caption{The frequency ($f$) of the tank-treading motion of a red blood cell in shear flow plotted against the shear rate ($\dot{\gamma}$). Simulated results using various models are denoted by filled symbols, while experimental data from~\citet{Fischer_2007} are represented by open circles. An exponential fit to the numerical results ($f\sim \dot{\gamma}^{0.92}$) is shown as a dashed line, and a linear fit ($f\sim \dot{\gamma}$) is depicted by the solid line.}
\label{fig:comp}
\end{center}
\end{figure}

Employing the same viscosity value for the suspending medium ($\eta_{ext} = 28.9$ mPa s) as reported in \citet{Fischer_2007}, which leads to $\lambda = 0.346$, we conducted a series of simulations using both single-layer and double-layer models to investigate the tank-treading frequency as a function of shear rate. Figure \ref{fig:comp} displays a comparison between our simulation results and experimental data. The results show that both single-layer and double-layer models indeed exhibit a power-law relationship with $\beta$ around 0.92. We note that our numerical results are also consistent with the results obtained by~\citet{peng_lipid_2013}, where they reported a value of approximately 0.91 for both their one-component and two-component models. In our simulations, we assumed a quasi-spherical reference shape with non-zero spontaneous curvature. We also explored scenarios with zero spontaneous curvature, resulting in negligible deviations from the primary results.

\section{Discussion}
\label{sec_Discussion}

\begin{figure}
\begin{center}
\includegraphics[width=3.5in]{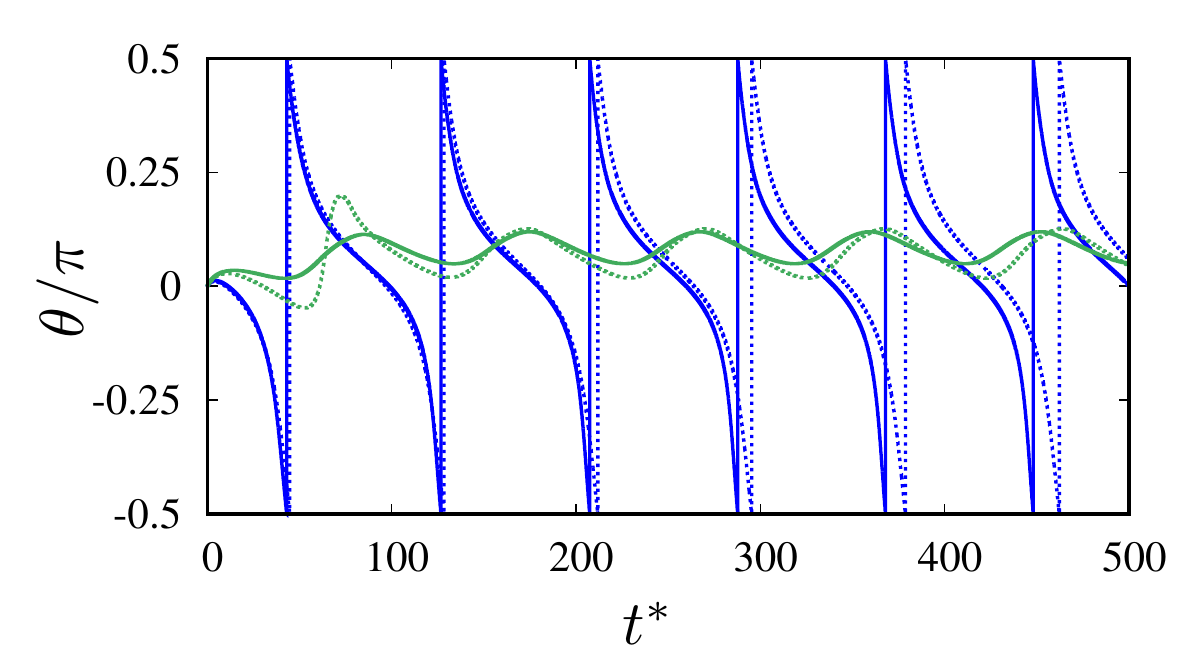}
\caption{
Time evolution of the inclination angle $\theta/\upi$ in the tumbling to tank-treading transition region at $[ \lambda = 0.417, Ca(\dot{\gamma},\mu_s) = 0.113 ]$, for capsule (\textcolor{blue}{blue}) and capsule-capsule (\textcolor{mygreen}{green}), using the quasi-spherical (solid line) and discocyte (dotted line) reference shapes for the cytoskeleton.}
\label{INT_Theta_RefShape}
\end{center}
\end{figure}

The choice of modelling strategy, either single or double layers, appears to have a significant impact on the dynamics of the RBC in shear flow. One way to assess the significance of this impact is to compare the influence of the cytoskeleton's reference shape. Figure \ref{INT_Theta_RefShape} provides such a comparison for the study point $[\lambda = 0.417, Ca(\dot{\gamma},\mu_s) = 0.113]$, located in the transition from tumbling to tank treading. The capsule and capsule-capsule models are compared with two alternatives of reference shape: discocyte or quasi-spherical (sphericity of $0.96$, which has already been used). For both modelling strategies, switching to the discocyte reference shape has a slight effect of reducing the oscillation frequency. Additionally, for the capsule-capsule model, a slight increase in the amplitude of the oscillations can be observed. However, for neither strategy does the influence go so far as to alter the nature of the dynamics. The capsule model remains in tumbling dynamics, while the capsule-capsule model remains in tank treading. In other words, distinguishing the cytoskeleton from the bilayer appears to have a greater impact.

\begin{figure}
\begin{center}
\includegraphics[width=5.3in]{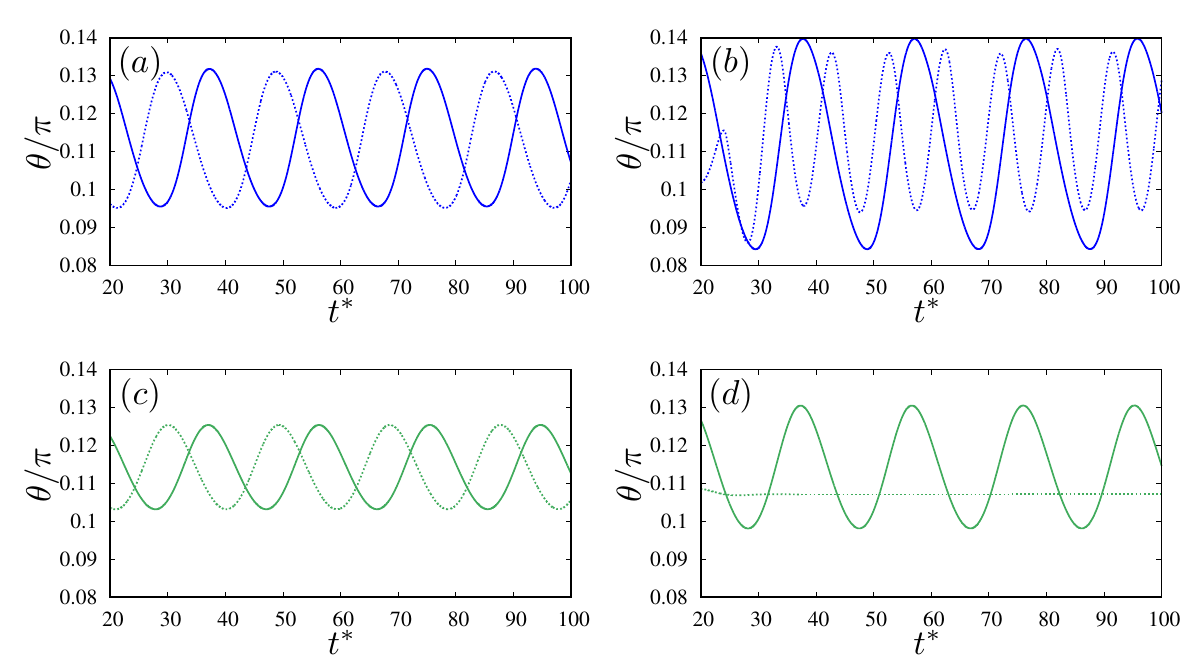}
\caption{
Impact of switching from the quasi-spherical (\textit{a},\textit{c}) to the discocyte (\textit{b},\textit{d}) reference shape on inclination angle in tank-treading regime
at $[ \lambda = 0.4, Ca(\dot{\gamma},\mu_s) = 0.5 ]$, 
for capsule (\textcolor{blue}{blue}) and capsule-capsule (\textcolor{mygreen}{green}).
The RBC's axis of symmetry is initially in the shear plane (solid line) or aligned with the axis of vorticity (dotted line).
}
\label{TT_Discocyte_Theta}
\end{center}
\end{figure}

\begin{figure}
\begin{center}
\includegraphics[width=5.3in]{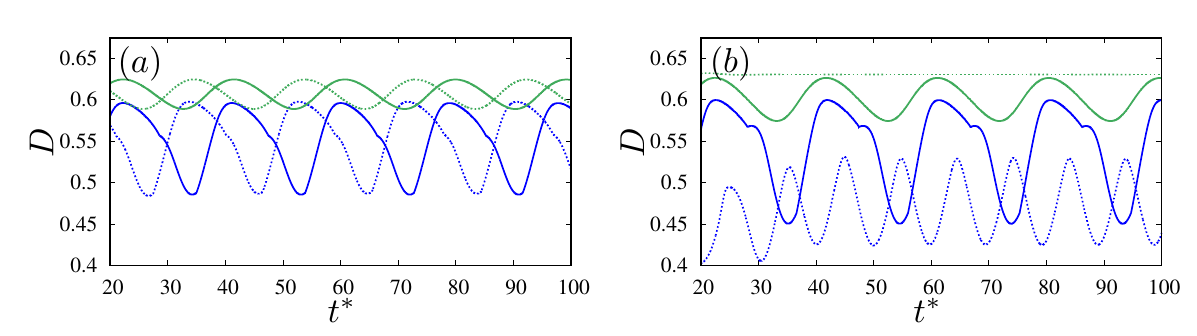}
\caption{
Impact of switching from the quasi-spherical (\textit{a}) to the discocyte (\textit{b}) reference shape on deformation in tank-treading regime at
$[ \lambda = 0.4, Ca(\dot{\gamma},\mu_s) = 0.5 ]$, 
for capsule (\textcolor{blue}{blue}) and capsule-capsule (\textcolor{mygreen}{green}).
The RBC's axis of symmetry is initially in the shear plane (solid line) or aligned with the axis of vorticity (dotted line).
}
\label{TT_Discocyte_D}
\end{center}
\end{figure}

\begin{figure}
\begin{center}
\includegraphics[width=5.3in]{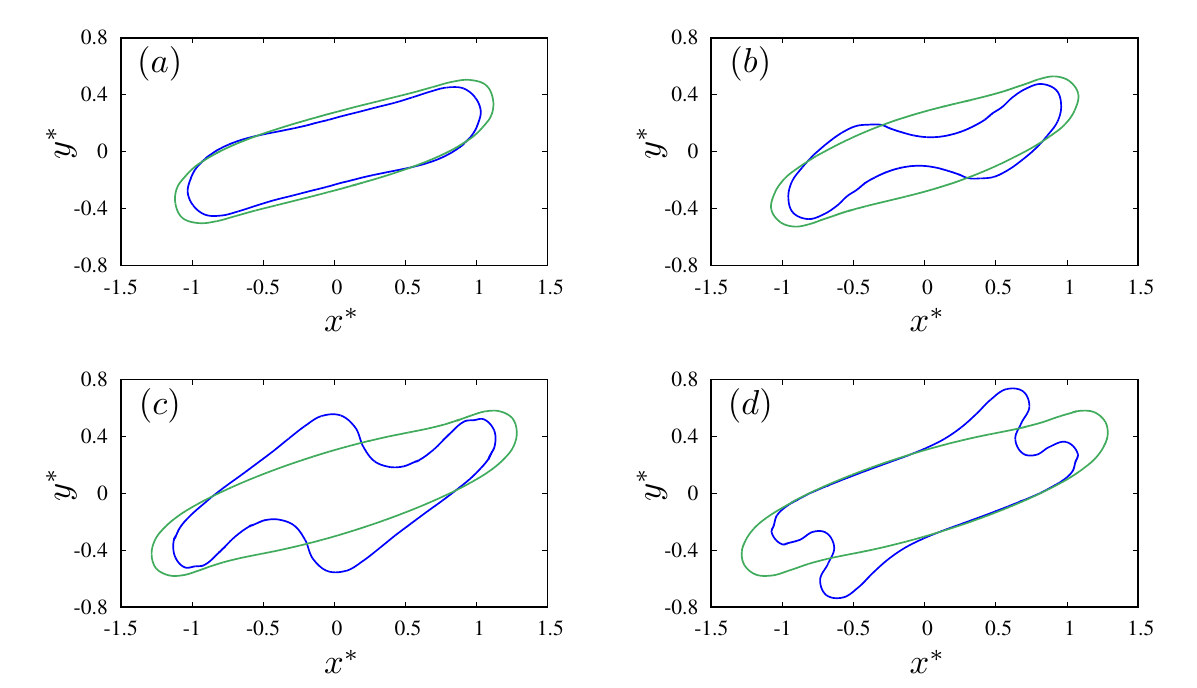}
\caption{Shape evolution [$(x^*,y^*)$ plane sectional drawing] with the discocyte reference shape on tank treading at $[ \lambda = 0.4, Ca(\dot{\gamma},\mu_s) = 0.5 ]$, for capsule (\textcolor{blue}{blue}) and capsule-capsule (\textcolor{mygreen}{green}), at $ t^* = 48 $ (\textit{a},\textit{c}), and $ t^* = 52 $ (\textit{b},\textit{d}). The RBC's axis of symmetry is initially in the shear plane (\textit{a},\textit{b}) or aligned with the axis of vorticity (\textit{c},\textit{d}).}
\label{TT_Discocyte_Shapes}
\end{center}
\end{figure}

Pushing the comparison in the tank-treading regime $[ \lambda = 0.4, Ca(\dot{\gamma},\mu_s) = 0.5 ]$, we observe a greater influence of the cytoskeleton's reference shape. The evolution curves of the inclination angle and the Taylor deformation parameter are shown in figures \ref{TT_Discocyte_Theta} and \ref{TT_Discocyte_D} for the two initial orientations of the RBC symmetry axis, i.e. in the shear plane or along its vorticity axis. Furthermore, the RBC shapes at $ t^* = 48 $ and $ t^* = 52 $ are compared in figure \ref{TT_Discocyte_Shapes}.
Switching to the discocyte reference shape has a significant impact, inducing undeniable discrimination between the two cases of the orientation of the RBC symmetry axis. This effect is evident for both the capsule and the capsule-capsule models, but it is more pronounced for the latter. When the axis of symmetry of the RBC is initially in the shear plane, an increase in the amplitude of the oscillations of the inclination angle and the deformation parameter is observed for both models. This result is consistent with the interpretation in terms of the energy barrier required to allow tank-treading movement, which is greater for the discocyte reference shape than for the quasi-spherical one. However, when the axis of symmetry of the RBC is aligned with the axis of vorticity, this energy barrier is reduced. For the capsule model, this results in a smaller oscillation amplitude for both the angle and the deformation, with a lower average deformation intensity. However, the most significant consequence is the doubling of the oscillation frequency. In contrast, for the capsule-capsule model, the barrier effect seems to disappear completely. The inclination angle and the deformation parameter take a constant value, and the RBC switches to pure tank-treading dynamics. However, the intensity of the deformation increases, unlike the capsule model, because of a greater influence of the dimple from the axis of symmetry towards the periphery for the more rigid capsule model. Here, the difference in modelling strategy between one or two layers still has a strong impact, but its influence combines with that of the reference shape for the cytoskeleton.

We conducted simulations with a non-zero spontaneous curvature for the quasi-spherical reference shape. However, we do not present the corresponding curves as they are very similar to those obtained with zero spontaneous curvature. 

We emphasise the significance of conducting a comparative study that minimises the discrepancies arising from specific numerical implementations, such as the method for solving flows and the geometric representation of surfaces. It is crucial for these numerical aspects to share the same level of precision. While the choice of our modelling may be subject to debate, including the numerical method and solution algorithms, our approach, built upon the relatively recent finite-element method known as isogeometric analysis (IGA), provides a coherent and consistent numerical framework. However, the dissipative contribution of the coupling between the cytoskeleton and the lipid bilayer is not present in single-layer models. Although our study does not consider relaxation dynamics, it is still legitimate to question the sensitivity of double-layer models to the friction coefficient $C_f$. To investigate this, we repeated simulations by varying $C_f$ by a factor of ten, but the impact was almost negligible. While it may seem surprising at first glance, the intensity of dissipation cannot be directly linked to the friction between the cytoskeleton and the lipid bilayer, since increasing the friction intensity decreases the velocity differential between the cytoskeleton and the lipid bilayer. 

\begin{figure}
\begin{center}
\includegraphics[width=5.4in]{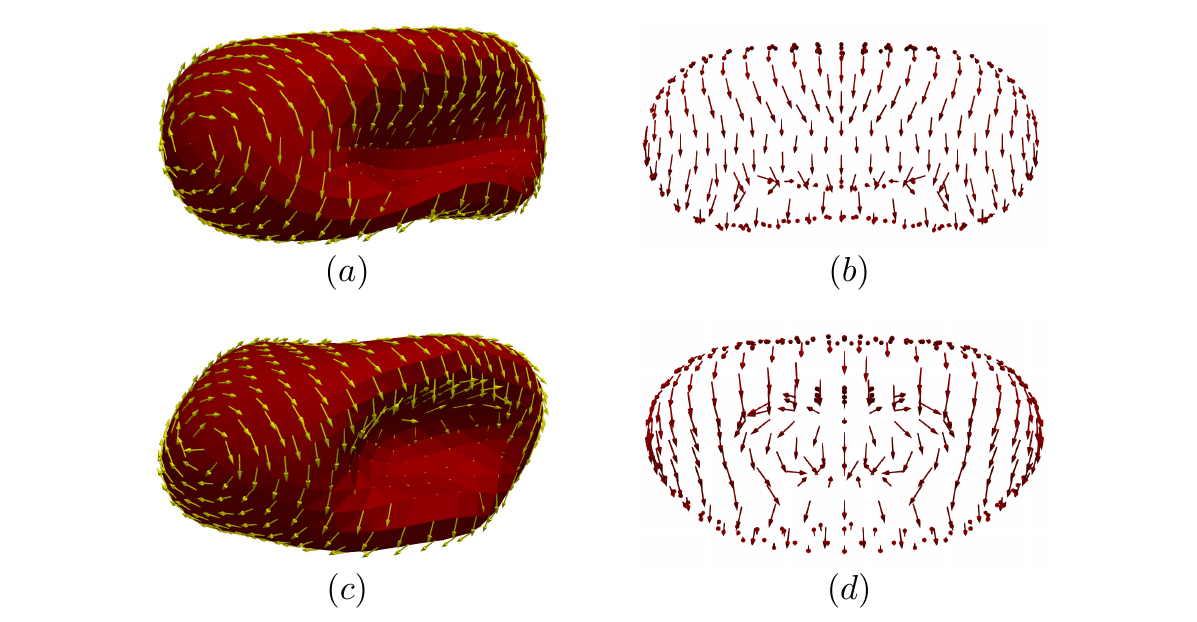}
\caption{Surface velocity fields in the multilobe region at $[ \lambda = 6.67, Ca(\dot{\gamma}, \mu_s) = 1.5 ]$, for the capsule-capsule (\textit{a}) and vesicle-capsule (\textit{c}) models, when $ D = D_\mathrm{min} $. (\textit{b},\textit{d}) The corresponding surface velocity field inside the concavity projected on a plane perpendicular to the concavity.}
\label{Multilobe_Velocity}
\end{center}
\end{figure}

The double-layer models used in our study clearly differ from the single-layer capsule model. However, beyond the surface incompressibility constraint that our study has revealed, what are the fundamental differences between the capsule-capsule and vesicle-capsule models, and why is it important to model the lipid bilayer as a fluid film? A fluid nature would allow the lipid bilayer to more easily adapt to ambient fluid flow, and even exhibit vortices in the surface velocity field if the energy balance is improved. In contrast, modelling the lipid bilayer as a solid shell would make such a scenario impossible. While tank-treading dynamics are compatible with a solid membrane and have been observed experimentally for capsules, it is unclear whether all the dynamics adopted by a red blood cell in a shear flow exhibit this behaviour. To investigate this further, we extended our comparison between capsule-capsule and vesicle-capsule models to multilobe dynamics $[\lambda = 6.67, Ca(\dot{\gamma}, \mu_s) = 1.5]$. Figure~\ref{Multilobe_Velocity} illustrates this comparison, showing comparable shapes obtained with both models when the Taylor parameter passes through its minimum. However, the vesicle-capsule model exhibits two counter-rotating vortices in the concavity, unlike the capsule-capsule model, which widens the concavity. This difference in surface flow topology may have a more significant effect on other aspects, including frictional dissipation between the cytoskeleton and lipid bilayer and the physiological role of the plasma membrane, which relies on its fluidity for proteins to diffuse widely and be easily mobilised to fulfil biological functions.

%%%%%%%%
\section{Conclusions}
\label{sec_Conclusion}

The red blood cell is a commonly studied area in the field of biomechanics, and it is traditionally modelled as a single-layer capsule or vesicle. While a few double-layer models exist, they often require two separate meshes -- one for the lipid bilayer and the other for the cytoskeleton. In this paper, we have presented an alternative double-layer membrane model that utilises a single mesh, significantly reducing computational complexity. Through computational assessment of different modelling strategies, including both single-layer and double-layer models, we have examined their respective effects on RBC dynamics and their potential implications in understanding the biomechanics of RBCs.

By analysing extensional flow, we have gained insight into the behaviour of the RBC membrane and why single-layer models like the capsule model and double-layer models like the capsule-capsule and vesicle-capsule models are not equivalent. The capsule model fails to distinguish the tangential kinematics of the cytoskeleton and the lipid bilayer, while in reality the latter only drives the former through the action of frictional forces of the lipids on the junction proteins. 
The cytoskeleton has a degree of freedom of tangential sliding, allowing it to relax its elastic stresses to an imposed surface shape, which is effective in double-layer modelling strategies but prohibited in the single-layer capsule model.  As a result, our findings show that the elastic strain energy of the cytoskeleton and the RBC membrane as a whole increase more slowly in the double-layer modelling strategies during extensional flow, resulting in a non-negligible increase in the elongation of the RBC. 

We aimed to investigate the influence of the interaction between the lipid bilayer and cytoskeleton layers of the RBC membrane on its dynamics in simple shear flow. To achieve this, we considered four points $[ \lambda, Ca(\dot{\gamma}, \mu_s)]$ of the phase diagram that corresponded to the dynamic regimes of tumbling, tank treading, transition to tank treading, and multilobe-shaped RBCs under very high shear. These points were chosen because they represent the richness of RBC behaviour in shear flow while keeping the RBC symmetry axis in the shear plane. For tank-treading dynamics, we also considered the alternative where the symmetry axis remains aligned with the vorticity axis. The modelling strategies were compared using the usual indicators of inclination angle and the Taylor deformation parameter, along with other indicators such as the shape of the RBC or the velocity field on its surface.

Our results show that modelling the RBC membrane as a single material surface or as two structures that can slide relative to each other is not equivalent. For all the study points, this difference in behaviour is always present, although it may be more or less marked depending on the nature of the dynamics considered. Our numerical methods were accurate, and any criticism of our choices would affect all RBC modelling strategies compared. The mechanical properties we considered were those recognised for a healthy RBC, but their assignment to one of the modelled components may vary depending on the strategy adopted. The sliding degree of freedom between the two layers in the double-layer models induced a dissipative phenomenon not considered in simpler models where the membrane is modelled as a block. However, we have shown that the sensitivity to the value of the surface friction coefficient is negligible, while the existence of this sliding degree of freedom is an important factor. Moreover, none of the cases we have considered involves relaxation dynamics.

Our study has yielded a surprising finding, emphasising the significance of the membrane modelling strategy for RBCs compared to the choice of the cytoskeleton's reference shape. Our results suggest that the combination of mechanical properties alone is insufficient, and careful consideration must be given to how these properties are incorporated. While the numerical efficiency favours a single-layer capsule approach, a double-layer model aligns more closely with biological reality. Fortunately, the additional computational cost of the double-layer model is minimal when utilising a single mesh, as demonstrated in our continuum-mechanical approach. Based on all the aspects considered, we conclude that the optimal sequence of RBC modelling strategies is as follows: double-layer vesicle-capsule and capsule-capsule models, followed by single-layer capsule and vesicle models if a particular model would be preferred.

However, an open question remains: is it necessary, useful, or avoidable to model the lipid bilayer as a vesicle rather than a capsule? To date, no such model has been widely adopted in the community. Our study identifies two mechanical properties contributed by the lipid bilayer: its surface incompressibility and its fluid nature. We show that the former consistently provides a marked stiffening effect compared to a quasi-incompressible capsule model, reducing up to 50\% of swinging oscillations in tank-treading dynamics. The importance of fluidity is less clear but becomes noticeable at very high shear rates, where contra-rotating vortices appear that cannot be reproduced with a capsule model. While the physiological significance of these complex flows within the plasma membrane remains uncertain, our study suggests that considering the fluidity of the lipid bilayer is still important. However, it is not the consideration of fluidity that is expensive in a vesicle model, but rather the rigorous consideration of the incompressibility constraint through a projection method in a space of surface divergence-free velocity fields. An alternative approach that would represent a good strategy and is already widely adopted in the vesicle community is to take into account the surface incompressibility constraint using a penalty method. 

Our initial motivation for conducting this study was rooted in numerical modelling concerns. However, our research has unexpectedly yielded insights that could advance our understanding of the biophysics of RBCs. One such insight is the unresolved question of the reference shape of the cytoskeleton. Our study on tank-treading dynamics has revealed a new phenomenon that could serve as an indicator in resolving this issue. Experimental research by \citet{minetti_dynamics_2019} suggests that in this regime, the RBC symmetry axis can be oriented in two ways -- aligned with the vorticity axis or in the shear plane. We have investigated both configurations, and our findings indicate that a reference shape close to a sphere results in indistinguishable steady-state dynamics for both orientations. However, with a discocyte reference shape, the effect is considerably different and accentuated in a double-layer model compared to a simple capsule model. The oscillation amplitude increases when the symmetry axis is in the shear plane and decreases when aligned with the vorticity axis. Notably, the oscillation frequency is halved in the latter case for the capsule model, whereas the capsule-capsule model eliminates the oscillations, leading to pure tank-treading dynamics. These observations suggest that the influence of reference shape on oscillation frequency is significant enough to warrant experimental verification.

Another point of interest is the fluidity of the lipid bilayer, which justifies the notion of a flow of its constituents and the associated friction on proteins linked to the cytoskeleton. Recent research has called into question the Singer and Nicolson model, particularly regarding the extent to which bilayer fluidity is constrained by the aggregation kinetics of its constituents and the corralling phenomenon \citep{kusumi_paradigm_2005,krapf_mechanisms_2015}. Surface viscosity and the possibility of vortices must be considered in this context. However, the limitation on the degree of sliding freedom of the cytoskeleton may not be immediately evident, since
it involves the sliding of the corrals themselves, which warrants further investigation.
   
%%%%%%%%
\section*{Acknowledgements}

%The authors would like to thank X for careful reading of the manuscript. 
The project leading to this publication has received funding from France 2030, the French Government program managed by the French National Research Agency (ANR-16-CONV-0001) and from Excellence Initiative of Aix-Marseille University - A*MIDEX.
Centre de Calcul Intensif d'Aix-Marseille is acknowledged for granting access to its high performance computing resources. 
\\
\\
\noindent {\bf Declaration of interests.} The authors report no conflict of interest.

\bibliographystyle{jfm}
\bibliography{biblio}

\end{document}